\documentclass{article}

\usepackage{hyperref}
\usepackage{amssymb,amsfonts,amsmath}
\usepackage{cite,enumerate,float}
\usepackage{color}
\usepackage{tikz}
\usetikzlibrary{arrows,snakes,backgrounds}
\usepackage{url}
\usepackage[vcentermath]{youngtab}

\usepackage{relsize}

\usepackage{enumitem,varwidth}

\def\be{\begin{eqnarray}}
\def\ee{\end{eqnarray}}
\def\nn{\nonumber}

\def\p{\partial}

\def\wc{weak composition\ }
\def\wcs{weak compositions\ }

\def\NS{\mathfrak{P}}

\definecolor{red}{rgb}{1,0,0}
\definecolor{orange}{rgb}{1,0.5,0}
\definecolor{violet}{rgb}{0.7,0,1}

\hoffset= - 1.0in         

\begin{document}

\title{\vspace{1.5cm}\bf
Integrable systems inspired by DAHA and DIM algebra:\\ type $C^\vee C$ versus type $A$
}

\author{
Liudmila Bishler$^{a,b,c,}$\footnote{bishlerlv@lebedev.ru},
Andrei Mironov$^{a,b,c,}$\footnote{mironov@lpi.ru,mironov@itep.ru},
Alexander Popolitov$^{b,c,d,}$\footnote{popolit@gmail.com}
}

\date{ }

\maketitle

\vspace{-6.5cm}

\begin{center}
  \hfill FIAN/TD-14/26\\
  \hfill ITEP/TH-29/26\\
  \hfill IITP/TH-26/26\\
  \hfill MIPT/TH-24/26\\
\end{center}

\vspace{4.5cm}

\begin{center}
$^a$ {\small {\it Lebedev Physics Institute, Moscow 119991, Russia}}\\
$^b$ {\small {\it NRC ``Kurchatov Institute", 123182, Moscow, Russia}}\\
$^c$ {\small {\it Institute for Information Transmission Problems, Moscow 127994, Russia}}\\
$^d$ {\small {\it MIPT, Dolgoprudny, 141701, Russia}}
\end{center}

\vspace{.1cm}

\begin{abstract}
The Ding-Iohara-Miki (DIM) algebra (quantum toroidal algebra of $\widehat{gl_1}$) is related to a wide class of quantum many-particle integrable systems, a typical one being the Ruijsenaars trigonometric system with eigenfunctions that are a triad formed by the Noumi-Shiraishi power series, the Macdonald polynomials, and the Baker-Akhiezer multivariable function. Other integrable systems of this type are obtained from the Ruijsenaars system by twisting. At the same time, the Ruijsenaars Hamiltonians are directly related to the Hamiltonians of another quantum integrable system, the Cherednik DAHA Hamiltonians of type $A$ (and their twisted versions in the twisted case), due to the correspondence between the DIM algebra and the spherical DAHA. The eigenfunctions of the DAHA Hamiltonians are non-symmetric Macdonald polynomials. Similarly, there is a class of integrable DAHA Hamiltonians of type $C^\vee C$, the spherical version of which, in turn, allows one to generate integrable Koornwinder Hamiltonians. The eigenfunctions of these two integrable systems are, respectively, non-symmetric and symmetric Koornwinder polynomials, which are our main interest in this paper. Here we consider the cases of both type $A$ and type $C^\vee C$ systems, since they are sufficiently similar, and point out important distinctions between them.
\end{abstract}

\bigskip

\newcommand\smallpar[1]{
  \noindent $\bullet$ \textbf{#1}
}

\newpage

\tableofcontents

\section{Introduction}

Many-body integrable systems are often given by commuting Hamiltonians that are elements of commutative subalgebras of some infinite-dimensional algebra (in the $n$-body representation). This algebra can be a Lie algebra \cite{MMMP1} as in the example of $W_{1+\infty}$ algebra \cite{Pope,Awata,KR1}, or an associative algebra \cite{MMMP2} as in the case of affine Yangian algebra \cite{Tsy,Proch}, when the corresponding integrable systems are the trigonometric Calogero-Sutherland systems and their generalizations: there are infinitely many commutative subalgebras in the affine Yangian algebra, and, hence, infinitely many corresponding integrable systems.

In fact, the affine Yangian algebra is a deformation of the $W_{1+\infty}$ algebra, and, in turn, can be obtained by a (rather intricate, \cite[sec.11]{MMP}) degeneration from the Ding-Iohara-Miki (DIM) algebra \cite{DI,Miki}, or equivalently, from the elliptic Hall algebra \cite{K,BS,S} (which is basically the same \cite{S,Feigin}). The DIM algebra is parameterized by 2 complex parameters $q,t$, and one of the commutative subalgebras of this algebra in the $n$-body representation is associated with the trigonometric Ruijsenaars system \cite{RS}, while all other can be obtained by $SL(2,\mathbb{Z})$ automorphisms of the algebra (Miki automorphisms \cite{Miki1}) \cite{MMP}. Hence, these automorphisms generate from the trigonometric Ruijsenaars-Schneider system a lot of new ones. In fact, the $SL(2,\mathbb{Z})$ algebra is generated by two transforms, one of them is rather trivial \cite{MMP}, while the other one generates non-trivial ``integer ray'' integrable systems \cite{MMP,MMP1}.

Now note that there is another class of integrable systems related to the trigonometric Ruijsenaars-Schneider systems. The relation is due to the equivalence of the DIM algebra and the spherical DAHA \cite{dFK1}. As a result, the Ruijsenaars Hamiltonians $H_k^{RS}$ can be realized as power sums of Cherednik Hamiltonians $C_i$ acting on the space of symmetric functions:
\be\label{DIMDAHA}
H_k^{RS}=\sum_{i=1}^nC_i^k\Big|_{symm}
\ee
$C_i$ being commutative elements of DAHA \cite{book:Ch-daha}. Similarly, the ``integer ray'' DIM Hamiltonians $\hat {\cal H}^{(a)}_k$ are obtained from $H_k^{RS}$ by applying $a$ times transformation ${\cal O}_h$ (with an additional simple rotation, see the detail in \cite{NSM4}) can be realized as power sums of $a$-twisted Cherednik Hamiltonians $\mathfrak{C}_i^{(a)}$:
\be\label{aDIMDAHA}
\hat {\cal H}^{(a)}_k=\sum_{i=1}^n \left({\mathfrak{C}}_i^{(a)}\right)^k\Big|_{symm}
\ee
$\mathfrak{C}_i^{(a)}$ also being commutative elements of DAHA \cite{MMP,NSM4}.

An essential issue when studying these integrable systems is constructing eigenfunctions of integrable Hamiltonians. In the case of Ruijsenaars Hamiltonians, these eigenfunctions are the Noumi-Shiraishi (NS) power series \cite{NS}, which are labelled by $n$ complex eigenvalues. At particular eigenvalues parameterized by the Young diagrams, the NS power series reduces to the Macdonald polynomials, and, at the particular value of $t=q^{-m}$, $m\in\mathbb{Z}_{>0}$, it reduces to the multivariable Baker-Akhiezer (BA) quasi-polynomial function due to O. Chalykh \cite{Cha}. The NS functions\footnote{In \cite{dFK2}, the NS function was called the universal solution.} along with their polynomial reductions, the Macdonald polynomials and the BA functions form {\bf a triad} \cite{MMP3}.

Eigenfunctions of the ``integer ray'' Hamiltonians $\hat {\cal H}^{(a)}_k$ are constructed only at $t=q^{-m}$, $m\in\mathbb{Z}_{>0}$: in this case, they are $a$-twisted Baker-Akhiezer functions \cite{CE,CF,MMP1}, while a counterpart of the NS series (universal solution in terms of \cite{dFK2}) is not known yet.

As for the Cherednik Hamiltonians $C_i$, their eigenfunctions are also known only at specific values of eigenvalues associated with weak compositions of the permutation group $S_n$ (see, however, some advance in \cite{NSM5}), i.e. of the Weyl group of the root system of type $A$. These eigenfunctions are non-symmetric Macdonald polynomials \cite{Opd95,Mac96,Che95}. A proper sum of these eigenfunctions gives rise to the symmetric Macdonald polynomials, i.e. the sum of eigenfunctions of the Cherednik Hamiltonians gives the eigenfunctions of the Ruijsenaars-Schneider (DIM algebra) Hamiltonians.

Similarly, the known eigenfunctions of the $a$-twisted Cherednik Hamiltonians ${\mathfrak{C}}_i$ are $a$-twisted non-symmetric Macdonald polynomials \cite{NSM1,NSM2,NSM4} also parameterized by the weak compositions of the permutation group $S_n$, and their proper sums give the symmetric eigenfunctions of the ``integer ray'' DIM Hamiltonians \cite{NSM3}.

This whole web of integrable systems is based on the correspondence between DIM algebra and spherical DAHA for the root system of type $A$ \cite{dFK1}. One also can look for a similar web that involves DAHA for other root systems. This is what we are doing in the present paper: we are discussing eigenfunctions of DAHA for the most general root system (apart from type $A$) of type $C^\vee C$ \cite{Macr} trying to stay in the description as close as possible to the type $A$ systems. Here an important point is that both the Macdonald polynomials \cite{Mac} (see also subsequent papers on the subject \cite{MacConj,CherednikConj,CherednikDAHA,Koorn}) and the BA functions \cite{Cha} are defined for any root system.

Nevertheless, we meet two essential problems. First of all, in the most general $C^\vee C$ case, an explicit counterpart of the NS power series (universal solution) is looking very involved, and we discuss a possible reason for this in sec.\ref{secbr}. Even in the rank one case of $C^\vee C_1$, a manifest expression for the BA function is not simple: the coefficients in front of $x$-monomials are not factorized. However, they do factorize for $B$, $C$, $B^\vee$ and $C^\vee$ root systems. Hence, in these four cases one can also obtain simpler expressions for the universal solutions. The second problem is that though in the $C^\vee C$ case, there is a counterpart of correspondence (\ref{DIMDAHA}) with the Ruijsenaars-Schneider Hamiltonians substituted by the Koornwinder-van Diejen Hamiltonians \cite{Koorn,vD}, and the space of symmetric functions substituted by functions invariant under permutations from the Weyl group of $C^\vee C_n$, however, twisted Hamiltonians in DAHA, or their counterpart ``integer ray'' Hamiltonians are not available. The reason is that DAHAs for other root systems do not have many automorphisms, which, in the $A$ type case, were in charge of the twisted Cherednik Hamiltonians. Because of the second problem, we do not discuss twisted systems in the $A$ type DAHA in this paper.

The paper is organized as follows. In section 2, we describe the eigenfunctions of the $A$ type Cherednik/Ruijsenaars-Schneider Hamiltonians, and, in the next sections, follow the same description for the $C^\vee C$ DAHA/Koornwinder-van Diejen Hamiltonians, however, in more details, and, hence, in more sections. That is, in section 3, we describe the $C^\vee C$ DAHA system and the Hamiltonians, both DAHA and Koornwinder-van Diejen ones. Then, in section 4, we describe the eigenfunctions: both non-symmetric and symmetric Koornwinder polynomials. In section 5, we consider a particular rank one case, when the Koornwinder polynomials become the Askey-Wilson (AW) ones and describe perspectives of the triad description. At last, section 6 contains some discussion and concluding remarks.

\paragraph{Notation.} We use the standard notation for the $q$-Pochhammer symbol:
\be
(x;q)_n:=\prod_{j=0}^{n-1}(1-xq^j),\ \ \ \ \ \ \ \ \ (x_1,x_2,\ldots,x_k;q)_n:=\prod_{i=1}^k(x_i;q)_n
\ee
The notation $x^y$ for $n$ variables $x_i$, $y_i$, $i=1,\ldots,n$ means
\be
x^y:=\prod_{i=1}^n x_i^{y_i}
\ee
and similarly $x^{\vec y}$.

For a sequence of $l$ integers $\{\alpha_1,\ldots,\alpha_l,0,\ldots,0\}$, $\alpha_l\ne 0$, we denote $|\alpha|:=\sum_{i=1}^l|\alpha_i|$, and the length of this sequence $l_\alpha$.

We define the quantum numbers
$$
[x]_q:={q^x-1\over q-1}
$$
and also
$$
\{x\}:=x-x^{-1}
$$

\section{Systems of type $A$}

\subsection{Hamiltonians}

\subsubsection{DAHA system of type $A$}

\paragraph{DAHA relations.} The $A_{n-1}$ DAHA is parameterized by 2 parameters $q$ and $t$ and consists of elements $T_i$ ($i=1,\ldots,n-1$), $X_j$, $C_j$ ($j=1,\ldots,n$) subject to the relations \cite{book:Ch-daha,dFK1}:

\bigskip

\begin{itemize}
\item At $i=1,\ldots,n-2$ (Hecke algebra):
\be
(T_i-1)(T_i+t^{-1})&=&0\nn\\
\phantom{.}[T_i,T_j]&=&0,\ \ \ \ \ \ \ |i-j|\ge 2\\
T_iT_{i+1}T_i&=&T_{i+1}T_iT_{i+1}
\ee
\item At $i=1,\ldots,n-1$:
\be\label{TCA}
tT_iC_{i+1}T_i&=&C_i\nn\\
\phantom{.}[T_i,C_j]&=&0\ \ \ \ \ \ i\ne j,j-1
\ee
\be\label{TXA}
tT_iX_iT_i&=&X_{i+1}\nn\\
\phantom{.}[T_i,X_j]&=&0\ \ \ \ \ \ i\ne j,j+1
\ee
\item At $i,j=1,\ldots,n$:
\be
\phantom{.}[C_i,C_j]=0\nn\\
\phantom{.}[X_i,X_j]=0
\ee
and, introducing,
\be
X:=X_1\ldots X_n,\ \ \ \ \ \ \ \ C:=C_1\ldots C_n
\ee
one also adds
\be
X_1C_2=tC_2T_1^2X_1\nn\\
CX_j=qX_jC\nn\\
XC_j=q^{-1}C_jX
\ee
\end{itemize}

\bigskip

\noindent
Now one can define
\be\label{pi}
\pi:=T_{n-1}^{-1}\ldots T_2^{-1}T_1^{-1}C_1
\ee
Then, at $i=1,\ldots,n-2$:
\be
\pi^{-1}T_i\pi=T_{i+1}
\ee
and
\be\label{Cpi}
C_i=t^{n-i}T_iT_{i+1}\ldots T_{n-1}\pi T_1^{-1}T_2^{-1}\ldots T_{i-1}^{-1}
\ee
We will also need an operator
\be\label{B}
B:=T_{n-1}\ldots T_2T_1 X_1
\ee
such that
\be\label{CB}
C_iB=B C_{i+1}
\ee

\paragraph{$x$-representation.}
This representation is defined to be
\be\label{xrep}
X_i F(x_1,x_2,\ldots,x_n):=x_iF(x_1,x_2,\ldots,x_n)
\ee
and the Hecke generators ($i=1,\ldots,n-1$) are realized in this representation as
\be\label{TxA}
T_i=1+{x_i-t^{-1}x_{i+1}\over x_i-x_{i+1}}(\sigma_{i,i+1}-1),\ \ \ \ \ i=1,\ldots,n-1
\ee
where $\sigma_{i,j}$ permutes $x_i$ and $x_j$.

The element $\pi$ acts in the $x$-representation as
\be\label{pix}
\pi F(x_1,x_2,\ldots,x_n)=F(qx_n,x_1,\ldots,x_{n-1})
\ee
and one can manifestly construct the Cherednik Hamiltonians $C_i$ in the $x$-representation using (\ref{TxA}), (\ref{pix}) and (\ref{Cpi}).

\subsubsection{Ruijsenaars Hamiltonians}

Hamiltonians of the trigonometric Ruijsenaars-Schneider system are \cite{RS}
\be\label{RH}
H_k^{RS}=\sum_{i_1<i_2<\ldots<i_k}{\left[\prod_{m=1}^kt^{\hat D_{i_m}}\Delta(x)\right]\over\Delta(x)}
\prod_{m=1}^k q^{\hat D_{i_m}}
\ee
where $\Delta(x)=\prod_{i<j}(x_i-x_j)$ is the Vandermonde determinant, and $\hat D_i:=x_i{\p\over\p x_i}$. Note that $H^{RS}_k=0$ at $k>n$ automatically. The first of the Hamiltonians is the celebrated Macdonald-Ruijsenaars operator
\be\label{MR}
H^{MR}=\sum_{i=1}^n\prod_{j\ne i}{tx_i-x_j\over x_i-x_j}q^{\hat D_i}
\ee
The Ruijsenaars-Schneider Hamiltonians, which are elements of a commutative subalgebra of the DIM algebra in the $n$-body representation \cite{MMP} can be constructed from the Cherednik Hamiltonians in the $x$-representation restricting the space of functions they are acting on:
\be
H_k^{RS}=\sum_{i=1}^nC_i^k\Big|_{symm}
\ee
where $\Big|_{symm}$ means the space of symmetric functions of $n$ variables $\{x_i\}$.
This reflects a correspondence between the DIM algebra and spherical DAHA \cite{dFK1}.

\subsection{Eigenfunctions}

\subsubsection{Eigenfunctions: generality}

The common polynomial eigenfunctions of $C_i$'s are enumerated by \wcs $\alpha$ (i.e. compositions admitting zero numbers, $\alpha_i\in\mathbb{Z}_{\ge 0}$) on which the symmetric group $S_n$ acts (i.e. the Weyl group for the root system of type $A$), and are called non-symmetric Macdonald polynomials $E_\alpha=E_{w\alpha^+}$, where $\alpha^+$ denotes the Young diagram, i.e. $\alpha^+_1\ge\alpha^+_2\ge\ldots\ge\alpha^+_n\ge 0$, and $w\in S_n$ is the shortest element that produces $\alpha$ from $\alpha^+$:
\be
C_i\cdot E_{\alpha}=\Lambda^{(i)}_{\alpha}\cdot E_{\alpha}\nn
\ee
The eigenvalues are
\be\label{ev}
\Lambda^{(i)}_{\alpha^+}=q^{\alpha^+_i}t^{n-i}
\ee
and
\be\label{evw}
\Lambda^{(i)}_{\alpha}=\Lambda^{(i)}_{w\alpha^+}=\Lambda^{w([1,n])_i}_{\alpha^+}=q^{\alpha_i}t^{w(\rho)_i}=
q^{\alpha_i}t^{n-1-\zeta_\alpha(i)}
\ee
where $\rho$ is the shifted Weyl vector with components $n-i$, and
\be
\zeta_\alpha(i):=\#\{k<i|\alpha_k\ge\alpha_i\}+\#\{k>i|\alpha_k>\alpha_i\}
\ee

The non-symmetric Macdonald polynomials are graded with the grading $|\alpha|:=\sum_{i=1}^n\alpha_i$. They have triangular expansions (hereafter we normalize the polynomials to have unit coefficient of the leading term):
\be\label{nsM}
E_{\alpha}=x^\alpha+\sum_{\beta<\alpha}C_{\alpha\beta}x^\beta
\ee
with the following ordering: for $\alpha=w(\alpha^+)$, $\beta=w'(\beta^+)$, $w,w'\in S_n$, one defines
$\alpha>\beta$ if $\alpha^+>\beta^+$ (e.g., in accordance with the lexicographic order), or, when $\alpha^+=\beta^+$, if the minimal length of $w$ is less than that of $w'$. This is called {\bf Bruhat order} \cite{HHL}.

\subsubsection{Constructing non-symmetric polynomials recursively\label{2.2.2}}

Now we describe an algorithmic procedure to construct the non-symmetric Macdonald polynomials (the so called Knop–Sahi recurrence) \cite{KS,HHL}. To this end, we note that the element $B$ (\ref{B}) acts on the non-symmetric Macdonald polynomials in the following way \cite{BF}:
\be\label{BE}
B\cdot E_{[\alpha_1,\ldots,\alpha_n]}(x_1,\ldots,x_n)=t^{-\#\{\alpha_i\le\alpha_1\}}
E_{[\alpha_2,\ldots,\alpha_{n},\alpha_1+1]}(x_1,\ldots,x_n)=q^{-\alpha_1}x_nE_{[\alpha_1\alpha_2,\ldots,\alpha_n]}(qx_{n},x_1,x_2,\ldots,x_{n-1})
\ee
and this allows one to lift the grading of $\alpha$ by one unit. After this, one can produce all $\alpha$ with the same $|\alpha|$ by acting on $\alpha^+$ by elements of the Weyl group $S_n$ generated by elements $s_i(\alpha)$, $i=1,\ldots,n-1$ such that
\be
s_i:(\alpha_1,\ldots,\alpha_{i-1},\alpha_i,\alpha_{i+1},\ldots,\alpha_n)&\to&
(\alpha_1,\ldots,\alpha_{i-1},\alpha_{i+1},\alpha_i,\ldots,\alpha_n)
\ee
using the action by elements $T_i$ (\ref{TxA}):
\be\label{perm}
\begin{array}{rcll}
T_iE_\alpha&=&E_\alpha,\ \ \ \ \ \ &\hbox{if}\ \ \ \ \ \alpha_i=\alpha_{i+1}\cr\cr
T_iE_\alpha&=&C^{(1)}_{i,\alpha}E_\alpha+E_{s_i\alpha},\ \ \ \ \ \ &\hbox{if}\ \ \ \ \ \alpha_i<\alpha_{i+1}\cr\cr
T_iE_\alpha&=&C^{(1)}_{i,\alpha}E_\alpha+C^{(2)}_{i,\alpha} E_{s_i\alpha},\ \ \ \ \ \ &\hbox{if}\ \ \ \ \ \alpha_i>\alpha_{i+1}
\end{array}
\ee
where $C^{(1)}_i$, $C^{(2)}_i$ are the rational functions of $q$ and $t$:
\be\label{C12}
C^{(1)}_{i,\alpha}:&=&-{(1-t)\Lambda^{(i,a)}_\alpha\over t(\Lambda^{(i,a)}_\alpha-\Lambda^{(i+1,a)}_\alpha)}=
-{(1-t)\Lambda^{(i)}_\alpha\over t(\Lambda^{(i)}_\alpha-\Lambda^{(i+1)}_\alpha)}=-{(1-t)\over t(1-r_{\alpha,i})}\\
C^{(2)}_{i,\alpha}:&=&{(\Lambda^{(i,a)}_\alpha-\Lambda^{(i+1,a)}_\alpha t)(\Lambda^{(i,a)}_\alpha-t^{-1}\Lambda^{(i+1,a)}_\alpha)\over t(\Lambda^{(i,a)}_\alpha-\Lambda^{(i+1,a)}_\alpha)^2}=
{(\Lambda^{(i)}_\alpha-\Lambda^{(i+1)}_\alpha t)(\Lambda^{(i)}_\alpha-t^{-1}\Lambda^{(i+1)}_\alpha)\over t(\Lambda^{(i)}_\alpha-\Lambda^{(i+1)}_\alpha)^2}={(1-tr_{\alpha,i})(1-t^{-1}r_{\alpha,i})\over t(1-r_{\alpha,i})^2}\nn
\ee
and we introduced the quantity
\be
r_{\alpha,i}:={\Lambda_\alpha^{(i+1)}\over\Lambda_\alpha^{(i)}}=q^{\alpha_{i+1}-\alpha_i}t^{\zeta(\alpha)_i-\zeta(\alpha)_{i+1}}\nn
\ee

For instance, in order to get the solution associated with \wc $[0,0,3]$, one starts the recursion with $E_{[0,0,0]}=1$ and uses the following sequence:
\be
[0,0,0]\stackrel{B}{\longrightarrow}[0,0,1]\stackrel{T_2}{\longrightarrow}[0,1,0]\stackrel{T_1}{\longrightarrow}[1,0,0]
\stackrel{B}{\longrightarrow}[0,0,2]\stackrel{T_2}{\longrightarrow}[0,2,0]\stackrel{T_1}{\longrightarrow}[2,0,0]
\stackrel{B}{\longrightarrow}[0,0,3]\nn
\ee

\subsubsection{Symmetric Macdonald polynomials}

While the eigenfunctions of the Cherednik Hamiltonians are the non-symmetric Macdonald polynomials enumerated by \wcs the eigenfunctions of the trigonometric Ruijsenaars-Schneider Hamiltonians are the symmetric Macdonald polynomials $M_\lambda$ enumerated by the dominant integral weights $\lambda$. They can be obtained \cite{CO} from the non-symmetric Macdonald polynomials by summing over the Weyl group $S_n$, i.e. over all permutations of the partition $\lambda$:
\be\label{nss}
M_{\lambda}=\sum_{{\alpha=w\cdot\lambda}\atop{w\in S_n}} E_\alpha\cdot\left(\prod_{(i,j):\ {{\alpha_j>\alpha_i}\atop{j>i}}}{\Lambda^{(i)}_\alpha-t^{-1}\Lambda^{(j)}_\alpha\over
\Lambda^{(i)}_\alpha-\Lambda^{(j)}_\alpha}\right)
\ee
where $\lambda$ is a partition (Young diagram), the product in the summand runs over pairs of $(i,j)$ such that $\alpha_i<\alpha_j$ at $i<j$, and the sum runs over all permutations $w$ from the symmetric group $S_n$. This gives the symmetric Macdonald polynomials in the standard normalization of the $P$ polynomials \cite{Macbook}, which corresponds to the unit coefficient in front of the leading term in the triangular expansion:
\be\label{sM}
M_{\lambda}=x^\lambda+\sum_{\beta<\alpha}{\cal C}_{\lambda\mu}x^\mu
\ee
Note that formula (\ref{nss}) can be rewritten a bit differently: first, one obtains $\beta$ from $\alpha^+$ by a minimal length signed permutation, $\sigma\in S_n$, which is a product of transpositions $\sigma=\prod_k s_{i_k}$, $i_k\in [1,\ldots,n]$. Then, one associates with each transposition $s_i$ acting on $\alpha$ a multiplier $N_\alpha(s_i)$:
\be\label{N}
N_\alpha(s_i)={\Lambda^{(i+1)}_\alpha-t^{-1}\Lambda^{(i)}_\alpha\over
\Lambda^{(i+1)}_\alpha-\Lambda^{(i)}_\alpha}={1\over t}{1-tr_{\alpha,i}\over 1-r_{\alpha,i}}
\ee
Now one obtains an equivalent presentation of (\ref{symm}) in the form
\be\label{NM}
M_{\lambda}=\sum_{{\alpha=w\cdot\lambda}\atop{w\in S_n}} E_\alpha\cdot\prod_k N_{\alpha^{(k)}}(s_{i_k})
\ee
where $\alpha{(k)}$ is the partition obtained after application of $k-1$ transpositions so that
  $\alpha^{(1)} = \alpha^+$, $\alpha^{(2)} = s_{i_1} \alpha^+$, and so on.

\subsubsection{Orthogonality}

There is a scalar product on the symmetric Macdonald polynomials (``another scalar product'') \cite{Macbook},
\be
\Big<f\Big|g\Big>_s&=&{1\over n!}\oint_0\prod_{i=1}^n{dx_i\over x_i}f(x)g(x^{-1})
\prod_{i\ne j}{(x_i/x_j;q)_\infty\over(tx_i/x_j;q)_\infty}
\ee
so that the polynomials are orthogonal w.r.t. this scalar product:
\be
\Big<M_\lambda\Big|M_\mu\Big>_s\sim\delta_{\lambda\mu}
\ee
This allows one to unambiguously obtain them using the triangular expansion (\ref{Ktr}).

A similar scalar product also exists for the non-symmetric Macdonald polynomials \cite{CheIP}:
\be\label{nsspM}
\Big<f\Big|g\Big>_{ns}&=&\Big<f,g\Big>_{ns}=\prod_{i=1}^n\oint{dx_i\over x_i}f(x_i;q,t)g(x_i^{-1};q^{-1},t^{-1})\prod_{i>j}{(x_i/x_j;q)_\infty(qx_j/x_i;q)_\infty
\over (tx_i/x_j;q)_\infty(tqx_j/x_i;q)_\infty}\nn\\
\Big<E_\alpha\Big|E_\beta\Big>_{ns}&\sim&\delta_{\alpha\beta}
\ee
It is technically less effective in evaluating the polynomials from the triangular expansion (\ref{sM}) and using this orthogonality relation as compared with the recursive construction of sec.\ref{2.2.2}, since the scalar product involves the polynomials both at $q,t$ and at their inverses.

\subsubsection{Evaluation}

The non-symmetric Macdonald polynomials are factorized at the point $x_i={1\over\Lambda_{\emptyset}^{(i)}}=t^{i-n}$ \cite{MacConj,CherednikConj}:
\be\label{EfM}
E_\alpha\left({1\over{\Lambda}^{(i)}_\emptyset}\right)=F_{\alpha^+}^{(0)}\prod_k F_\alpha(s_{i_k})
\ee
Here one has to obtain $\alpha$ from $\alpha^+$ by a minimal length permutation, $\sigma\in S_n$, which is a product of transpositions $\sigma=\prod_k s_{i_k}$, $i_k\in [1,\ldots,n]$, and then, one associates with each transposition a multiplier $F_\alpha(s_i)$ (note the striking similarity with (\ref{N})):
\be
F_\alpha(s_i)={1-r_{\alpha,i}\over 1-t^{-1}r_{\alpha,i}}
\ee
while
\be\label{F+}
F_{\lambda}^{(0)}=t^{\sum_i(i-n)\lambda_i}\cdot
\prod_{i<j}{(qt^{j-i+1};q)_{\lambda_i-\lambda_j}\over(qt^{j-i};q)_{\lambda_i-\lambda_j}}
\ee

The symmetric Macdonald polynomials are factorized at the same point $x_i={1\over\Lambda_{\emptyset}^{(i)}}=t^{i-n}$ \cite{MacConj,CheF}:
\be
M_\lambda\left({1\over\Lambda_{\emptyset}^{(i)}}\right)=t^{\sum_ii\lambda_i}\cdot
\prod_{i<j}{(t^{j-i+1};q)_{\lambda_i-\lambda_j}\over(t^{j-i};q)_{\lambda_i-\lambda_j}}
\ee
where $\nu_\lambda:=2\sum_i(i-1)\lambda_i$. This formula differs from (\ref{F+}) only by additional factors of $q$ in the $q$-Pochhammer symbols.

\subsubsection{Duality relations\label{sdua}}

Among the Macdonald polynomials, there is a duality, which is a direct corollary of the Ruijsenaars duality \cite{Rdu,Etin,MMZ,MMdell}.
For non-symmetric Macdonald polynomials, it is of the form \cite{CherednikConj}
\be \label{EEM}
{E_\alpha\left({1\over{\Lambda}^{(i)}_\beta}\right)\over
E_\alpha\left({1\over{\Lambda}^{(i)}_\emptyset}\right)}={{E}_\beta\left({1\over\Lambda^{(i)}_\alpha}\right)\over
{E}_\beta\left({1\over\Lambda^{(i)}_\emptyset}\right)}
\ee
while, for the symmetric Macdonald polynomials, it looks like \cite{Macbook}
\be\label{Mdua}
{M_{\alpha^+}\left({\Lambda}^{(i)}_{\beta^+}\right)\over
M_{\alpha^+}\left({\Lambda}^{(i)}_{\emptyset}\right)}={{M}_{\beta^+}\left(\Lambda^{(i)}_{\alpha^+}\right)\over
{M}_{\beta^+}\left(\Lambda^{(i)}_{\emptyset}\right)}
\ee
Because of the symmetry of the Macdonald polynomials $M$, one can write this equality at inverse eigenvalues similarly to (\ref{EEM}).

Note that these are exactly the quantities entering the r.h.s. of the CMM formulas \cite{CheCMM,EK,CE,MMPU}.

\subsubsection{Stability}

The stability property of the non-symmetric Macdonald polynomials looks as follows
\be\label{stab}
\begin{array}{ll}
E_{[\alpha_1,\alpha_2,\ldots,\alpha_{n-1},\alpha_n]}(x_1,x_2,\ldots,x_{n-1},x_n)\Big|_{x_n=0}
=E_{[\alpha_1,\alpha_2,\ldots,\alpha_{n-1}]}(x_1,x_2,\ldots,x_{n-1})&\ \ \ \ \ \hbox{if}\ \ \ \alpha_n=0\cr
E_{[\alpha_1,\alpha_2,\ldots,\alpha_{n-1},\alpha_n]}(x_1,x_2,\ldots,x_{n-1},x_n)\Big|_{x_n=0}=0&\ \ \ \ \ \hbox{if}\ \ \ \alpha_n\ne 0
\end{array}
\ee

\subsection{Triad of type $A$}

\subsubsection{Symmetric polynomials: branching rules}

One can always expand the Macdonald polynomial of $n$ variables into that of $n-1$, and it turns out that the coefficients of this expansion are factorized \cite{Macbook}:
\be\label{br}
M_\mu(x_1,\ldots,x_n;q,t)=\sum_\nu M_{\mu/\nu}(1;q,t)x_1^{|\mu|-|\nu|}M_\nu(x_2,\ldots,x_{n};q,t)
\ee
\be\label{brc}
M_{\mu/\nu}(1;q,t)=\prod_{1\le i<j\le n}{(q^{\nu_i-\mu_j+1}t^{j-i-1};q)_{\mu_i-\nu_i}\over (q^{\nu_i-\mu_j}t^{j-i};q)_{\mu_i-\nu_i}}
\prod_{1\le i\le j\le n-1}{(q^{\nu_i-\nu_j}t^{j-i+1};q)_{\mu_i-\nu_i}\over(q^{\nu_i-\nu_j+1}t^{j-i};q)_{\mu_i-\nu_i}}
\ee
This branching rule is easily extendable to arbitrary complex values of $\{\mu_i\}$, i.e. of lengths of the Young diagram lines.

\subsubsection{$A_1$ triad}

\paragraph{Noumi-Shiraishi function.} We start from the simplest example of $n=2$. In this case,
\be\label{M2}
M_{[\mu_1,\mu_2]}=x_1^{\mu_1}x_2^{\mu_2}\sum_{k=0}^{\mu_1-\mu_2}\left({x_2\over x_1}\right)^k\left({q\over t}\right)^k
{(q^{\mu_2-\mu_1},t;q)_{k}\over (q^{\mu_2-\mu_1+1}t^{-1},q;q)_{k}}
\ee
First, one can lift the upper limit of summation, since it is automatically ensured by the $q$-Pochhammer coefficients.
Then, one can put $\mu_{1,2}$ here to be arbitrary complex numbers, and write (\ref{M2}) as the power series
\be\label{NS2}
\NS_{q,t}(\vec x,\vec y)=x_1^{\lambda_1}x_2^{\lambda_2}\sum_{k=0}^{\infty}\left({x_2\over x_1}\right)^{k+{1\over 2}\log_q t}
{\Big({qy_1\over ty_2}q^k,t;q\Big)_{k}
\over \Big({y_1\over y_2}q^k,q;q\Big)_{k}}
\ee
where, for future convenience, we introduced the quantities $y_1:=q^{\lambda_1}:=q^{\mu_1}t^{{1\over 2}}$, $y_2:=q^{\lambda_2}:=q^{\mu_2}t^{-{1\over 2}}$. This power series is an eigenfunction of the Macdonald-Ruijsenaars Hamiltonian (\ref{MR}), it is called Noumi-Shiraishi (NS) function \cite{NS}, and has {\bf two} polynomial reductions.

\paragraph{Polynomial reductions.} The first reduction is at integer $\mu_{1,2}$, $\mu_1\ge\mu_2\ge 0$, and it returns us to the Macdonald polynomial of two variables.

The other (quasi)polynomial reduction is at $t=q^{-m}$ with integer non-negative $m$. In this case, the infinite series in (\ref{NS2}) also becomes a finite sum with $m+1$ terms, and this is the Baker-Akhiezer (BA) function introduced by O. Chalykh \cite{Cha}:
\be
\Psi_m(\vec x,\vec y;q)=x_1^{\lambda_1}x_2^{\lambda_2}\sum_{k=0}^{m}\left({x_2\over x_1}\right)^{k-{m\over 2}}
{\Big({y_1\over y_2}q^{k+m+1},q^{-m};q\Big)_{k}
\over \Big({y_1\over y_2}q^k,q;q\Big)_{k}}
\ee
Upon removing a trivial monomial factor of $(x_1x_2)^{\lambda_1+\lambda_2}$, one arrives \cite{MMP1} at
\be\label{BA2}
\overline{\Psi}_m(\vec x,\vec y;q)=x^{\lambda \over 2}\sum_{k=0}^{m}x^{{m\over 2}-k}
{\Big(yq^{k+m+1},q^{-m};q\Big)_{k}
\over \Big(yq^k,q;q\Big)_{k}}
\ee
where $x:={x_1\over x_2}$ and $y:={y_1\over y_2}=q^\lambda$.

Closing the triad, the two polynomial reductions are related by symmetrization over the Weyl group.

\subsubsection{$A_n$ triad}

This construction can be straightforwardly extended to arbitrary $n$: one can put $\mu_i$ arbitrary complex numbers and repeatedly apply the branching rule (\ref{br}):
\be\label{MNS}
M_\mu(x_1,\ldots,x_n;q,t)=\sum_{\{\nu_i\}} M_{\mu/\nu^{(1)}}(1;q,t)M_{\nu^{(1)}/\nu^{(2)}}(1;q,t)\ldots
M_{\nu^{(n-1)}/\nu^{(n)}}(1;q,t)x_1^{|\mu|-|\nu_1|}x_2^{|\nu_1|-|\nu_2|}\ldots x_n^{|\nu_{n-1}|-|\nu_n|}\nn\\
\ee
This nested anzatz \cite{MMP7} gives rise to the NS power series \cite{NS}, which is defined to be
\be\label{NS}
\NS_{q,t}(\vec x,\vec y)= x^{\lambda-\log_q t\cdot\rho}\cdot
\mathlarger{\mathlarger{\sum}}_{k_{ij}}\psi(\vec y,k_{ij};q,t)\prod_{1\le i<j\le N}\left({x_j\over x_i}\right)^{k_{ij}}
\ee
where again $y_i=q^{\lambda_i}$, the sum goes over all non-negative integer $k_{ij}$ with $i<j$,
and $\rho$ is the Weyl vector, i.e. $\rho_i={1\over 2}(n-2i+1)$. The coefficients $\psi(\vec y,k_{ij};q,t)$ are
\be
\psi(\vec y,k_{ij};q,t):&=&\prod_{n=2}^N\prod_{1\le i<n}{(t;q)_{k_{in}}
\over (q;q)_{k_{in}}}
{\Big({qy_i\over ty_n}q^{k_{in}-\sum_{a>n}(k_{ia}-k_{na})};q\Big)_{k_{in}}
\over \Big({y_i\over y_n}q^{k_{in}-\sum_{a>n}(k_{ia}-k_{na})};q\Big)_{k_{in}}}
\times\\
&\times&
\prod_{n=2}^N\prod_{1\le i<j< n}{\Big({qy_i\over ty_j}q^{k_{in}-\sum_{a>n}(k_{ia}-k_{ja})};q\Big)_{k_{in}}
\over \Big({y_i\over y_j}q^{k_{in}-\sum_{a>n}(k_{ia}-k_{ja})};q\Big)_{k_{in}}}
{\Big({ty_i\over y_j}q^{k_{in}+k_{jn}-\sum_{a>n}(k_{ia}-k_{ja})};q\Big)_{k_{in}}
\over
\Big({qy_i\over y_j}q^{k_{in}+k_{jn}-\sum_{a>n}(k_{ia}-k_{ja})};q\Big)_{k_{in}}}\nn
\label{c}
\ee
The NS functions are eigenfunctions of the Ruijsenaars-Schneider Hamiltonians (\ref{RH}):
\be
\hat H_k^{RS}\cdot\NS_{q,t}(\vec x,\vec y)=\left(\sum_iy_i^k\right)\cdot\NS_{q,t}(\vec x,\vec y)
\ee

\paragraph{Polynomial reductions.} There are again two polynomial reductions: choosing $y_i(\mu)=q^{\mu_i}t^{\rho_i}$ with $\{\mu_i\}$ being a partition, one obtains the Macdonald polynomial $M_\mu(x)$, while choosing $t=q^{-m}$, one generates the BA function $\Psi_m(\vec x,\vec y;q)$, and the sum of the BA functions over all permutations (over the Weyl group $S_n$) at $\vec y=\vec y(\mu)$ gives rise to the Macdonald polynomial:
\be\label{NSM}
M_\mu(x;q,t)&=&\NS_{q,t}(\vec x,\vec y(\mu))
\ee
\be\label{NSBA}
\Psi_m(\vec x,\vec y;q)&=&\NS_{q,q^{-m}}(\vec x,\vec y)
\ee
\be\label{BAM}
M_\mu(x;q,q^{-m})&=&\sum_{w\in S_n}\Psi_m(w\vec x,\{q^{\mu_i}t^{\rho_i}\};q)
\ee
Note that, in variance with the Macdonald reduction which is a symmetric polynomial in $x_i$, the BA function is not symmetric. However, we note that one can get a symmetric sum under one reduction condition, and obtain {\bf the same sum} under another eduction condition {\bf plus} summation over permutations. This is due to non-permutability of the two polynomial reductions \cite[sec.2.5]{MMP3}, which seems to be an essential feature of algebraic integrability setups \cite{Cha} and of $A_n$ triad structure in particular. 

\paragraph{BA function.} Crucially, the BA function (\ref{NSBA}) has an unambiguous definition up to a normalization factor, which has nothing to do with triad. Namely, consider a set of periodicity conditions \cite{Cha,MMP1}
\be\label{symm1}
\Psi_m(x_kq^j,\vec y;q)=\Psi_m(x_lq^j,\vec y;q)\ \ \ \ \  \forall k,l\ \ \hbox{and}\ \ 1\le j\le m\ \ \ \ \ \hbox{at}\ \ x_k=x_l
\ee
imposed on the (quasi)polynomial
\be
\Psi_m(\vec x,\vec y;q)= x^{\lambda-\log_q t\cdot\rho}\cdot
\mathlarger{\mathlarger{\sum}}_{k_{ij}}\psi(\vec y,k_{ij};q,t)\prod_{1\le i<j\le N}\left({x_j\over x_i}\right)^{k_{ij}}
\ee
Such defined BA function is unique. After multiplication by the normalization factor
\be
{\cal N}=\prod_{k>l}\prod_{j=1}^m\Big(\sqrt{y_k\over y_l}q^{-j}-\sqrt{y_l\over y_k}\Big)
\ee
$\Psi_m(\vec x,\vec y;q)$ becomes symmetric under the permutation of sets $\{x_i\}$ and $\{y_i\}$.

\section{Systems of type $C^\vee C$: Hamiltonians}

The very detailed construction of systems of type $C^\vee C$ is mostly due to \cite{Noumi,Sahi,Sahi2}, see also later discussions in \cite{St,Chalykh,CR}. We discuss this case maximally parallel to the $A$-series case of the previous section. However, in order to make some formulas simpler, hereafter, we use rather parameters $q^2$, $t^2$ instead of $q$, $t$.

\subsection{DAHA system of type $C^\vee C$}

\subsubsection{Set of DAHA relations}

The DAHA associated with root system $C^\vee C_n$ is defined by the relations \cite{Chalykh}
\begin{itemize}
\item Braid group relations:

\noindent
at $i,j=0,\ldots,n$:
\be
\phantom{.}[T_i,T_j]&=&0,\ \ \ \ \ \ \ |i-j|\ge 2
\ee
and, at $i=1,\ldots,n-2$:
\be
T_iT_{i+1}T_i&=&T_{i+1}T_iT_{i+1}
\ee
\item At $i=1,\ldots,n-1$ (Hecke relations):
\be
(T_i-t^2)(T_i+1)&=&0
\ee
and
\be
(T_0-ac/q^2)(T_0-1)&=&0\nn\\
(T_n-bd)(T_n-1)&=&0
\ee
\item At $i=0$, or at $i=n-1$ (reflection equation):
\be\label{TT}
T_iT_{i+1}T_iT_{i+1}&=&T_{i+1}T_iT_{i+1}T_i
\ee
\item At $i=1,\ldots,n-1$:
\be\label{TC}
T_iC_{i+1}T_i&=&t^2C_i\nn\\
\phantom{.}[T_i,C_j]&=&0\ \ \ \ \ \ i\ne j,j-1
\ee
\be\label{TX}
T_iX_iT_i&=&t^2X_{i+1}\nn\\
\phantom{.}[T_i,X_j]&=&0\ \ \ \ \ \ i\ne j,j+1
\ee
and also
\be
\phantom{.}[T_n,X_{n-1}]&=&0\nn\\
X_1^{-1}T_0+{ac\over q^4}T_0^{-1}X_1&=&{a+c\over q^2}\nn\\
T_nX_n+bdX_n^{-1}T_n^{-1}&=&b+d\nn\\
C_n^{-1}T_n+bdT_n^{-1}C_n&=&{db\over\eta}\Big(1+{ac\over q^2}\Big)
\ee
where $\eta:={\sqrt{abcd}\over q}$.
\item At $i,j=1,\ldots,n$:
\be
\phantom{.}[C_i,C_j]=0\nn\\
\phantom{.}[X_i,X_j]=0
\ee
\end{itemize}
The commutating Hamiltonians are given by (the normalization is chosen for future convenience)
\be
C_i={t^{2(i-n)}\over\eta}T_i\ldots T_{n-1}T_nT_{n-1}\ldots T_0T_1^{-1}\ldots T_{i-1}^{-1},\ \ \ \ \ \ \ i=1,\ldots,n
\ee

\subsubsection{$x$-representation (Noumi representation \cite{Noumi})}

In this representation, the Hecke generators ($i=1,\ldots,n-1$) are realized as follows
\be
T_0={ac\over q^2}-{(x_1-a)(x_1-c)\over (x_1^2-q^2)}\Big(\sigma_0q^{2\hat D_1}-1\Big)\nn\\
T_i=t^2+{(t^2x_i-x_{i+1})\over(x_i-x_{i+1})}(\sigma_i-1)\nn\\
T_n=bd-{(1-bx_n)(1-dx_n)\over(1-x_n^2)}(\sigma_n-1)
\ee
where $\sigma_{0}$ and $\sigma_{n}$ make inversions $x_{0}\to x_{0}^{-1}$ and $x_{n}\to x_{n}^{-1}$ accordingly.

\subsubsection{Parameters}

Thus, the system is parameterized by the six Koornwinder parameters $q,t,a,b,c,d$, which can also be implemented through six other standardly used parameters $q,t,t_0,t_n,u_0,u_n$.
The identification of parameters is as follows (any parameters can be permuted)
\be\label{abcdt}
a=-q\sqrt{t_0\over u_0},\ \ \ \ \ \ \ \ \ \
b=-\sqrt{t_n\over u_n},\ \ \ \ \ \ \ \ \ \
c=q\sqrt{t_0u_0},\ \ \ \ \ \ \ \ \ \
d=\sqrt{t_nu_n}
\ee
This set of parameters also has a set of dual parameters associated with it. The set of dual parameters is just $u_n,u_0,t_n,t_0$ instead of $t_0,u_0,t_n,u_n$, i.e. is formed by the permutation of parameters $t_0$ and $u_n$. Hereafter, we denote the replacement of parameters with their duals using a bar. In terms of the Koornwinder parameters $a,b,c,d$ this transformation is more involved, and describes the transition to the dual Askey-Wilson variables, \cite{AW}:
\be\label{dualAW}
\bar a={ad\over\eta},\ \ \ \ \ \ \ \ \ \ \bar b={bd\over\eta},\ \ \ \ \ \ \ \ \ \ \bar c={cd\over\eta},\ \ \ \ \ \ \ \ \ \
\bar d=\eta
\ee

\subsubsection{Root systems}

Generic values of parameters $a,b,c,d$ correspond to affine root system of type $(C^\vee C_n)$, while other affine root systems are associated with the following peculiar choices of the parameters (one can definitely permute $a$ and $c$, and $b$ and $d$, or just signs in front of them):

\bigskip

\bigskip

\noindent {\bf Reduced systems of roots:}
\hspace{2cm}
\begin{varwidth}{\linewidth}
\begin{itemize}[nosep]
\item[$B_n$] $(a,b,c,d)=(-1,-q,t_s,q)$
\item[$C_n$] $(a,b,c,d)=(-t_l,-qt_l,t_l,qt_l)$
\item[$D_n$] $(a,b,c,d)=(-q,-1,q,1)$
\item[$B_n^\vee$] $(a,b,c,d)=(-q,-t_l,q,t_l)$
\item[$C_n^\vee$] $(a,b,c,d)=(-1,-t_s,qt_s,q)$
\item[$BC_n$] $(a,b,c,d)=(-1,-qt_l,t_s,qt_l)$
\end{itemize}
\end{varwidth}

\bigskip

\bigskip

\noindent {\bf Non-reduced systems of roots:}
\hspace{2cm}
\begin{varwidth}{\linewidth}
\begin{itemize}[nosep]
\item[$(BC_n,C_n)$] $(a,b,c,d)=(-t_s,-qt_l,t_l,qt_l)$
\item[$(C_n^\vee,BC_n)$] $(a,b,c,d)=(-t_s,-qt_s,q,t_l)$
\item[$(B_n^\vee,B_n)$] $(a,b,c,d)=(-q,-t_s,q,t_l)$
\end{itemize}
\end{varwidth}

\bigskip

\bigskip

\noindent
Here $t_l$ ($t_s$) denotes $t_\alpha$ associated with longest (shortest) roots of the corresponding root systems. These formulas demonstrate the well-known fact that the Macdonald polynomials for root system $B_n$ (analogously, $C_n$) at $t_s=1$ ($t_l=1$) coincide with those for the root system $D_n$.

One immediately see that at $t_{l,s}=1$ the first four reduced systems of roots coincide, and they also coincide with the last non-reduced system: DAHA$_{B_n}=$DAHA$_{C_n}=$DAHA$_{D_n}=$DAHA$_{B_n^\vee}=$DAHA$_{(B_n^\vee,B_n)}$. This, in particular, means that coincide also the corresponding non-symmetric and symmetric polynomials.

\subsection{Koornwinder Hamiltonians}

\subsubsection{Rank 1 Hamiltonian}

The Koornwinder Hamiltonian is as follows (note that parameters $q$ and $t$ are squared as compared with the original Koornwinder notation):
\be\label{KH}
H^K_1:=\sum_{i=1}^nt^{2(1-n)}{(1-ax_i)(1-bx_i)(1-cx_i)(1-dx_i)\over \eta(1-x_i^2)(1-q^2x_i^2)}\prod_{j\ne i}
{(1-t^2x_ix_j)(1-t^2x_i/x_j)\over (1-x_ix_j)(1-x_i/x_j)}\left(q^{2\hat D_i}-1\right)+\nn\\
+\sum_{i=1}^nt^{2(1-n)}{(1-ax_i^{-1})(1-bx_i^{-1})(1-cx_i^{-1})(1-dx_i^{-1})\over \eta(1-x_i^{-2})(1-q^2x_i^{-2})}\prod_{j\ne i}
{(1-t^2x_i^{-1}x_j^{-1})(1-t^2x_j/x_i)\over (1-x_i^{-1}x_j^{-1})(1-x_j/x_i)}\left(q^{-2\hat D_i}-1\right)
\ee
with the eigenvalues
\be
\Lambda_\lambda=\sum_{j=1}^n\{\eta t^{2(n-j)j}q^{\lambda_j}\}\{q^{\lambda_j}\}
\ee
This Hamiltonian coincides with that obtained from the Weyl-symmetric combination of the Cherednik Hamiltonians, when acting on the Weyl-symmetric functions:
\be
H^K_1=\sum_{i=1}^n(C_i+C_i^{-1})-[n]_t\Big(t^{n-1}\eta+{1\over t^{n-1}\eta}\Big)\Big|_{symm}
\ee

\subsubsection{Commutative van Diejen-Koornwinder Hamiltonians}

Since the system is integrable, there are in total $n$ commuting Hamiltonians, their explicit form being \cite{vD}
\begin{equation}
    \begin{aligned}
        H_r^{K} = \sum_{\substack{J \subset \{1,\dots,n\},\, |J| = r \\ \varepsilon_j = \pm 1,\, j\in J}} \,\, \sum_{\substack{\emptyset \subsetneqq J_1 \subsetneqq \dots \subsetneqq J_s = J \\ 1 \leq s \leq r}} (-1)^{s-1} \prod_{1\leq s'\leq s} h_{\varepsilon(J_{s'} \setminus J_{s'-1});J_{s'}^C} (q^{2\sum_{j\in J_1}\varepsilon_j \hat D_j } -1)
    \end{aligned}
\end{equation}
The first sum runs over all subsets of $\{1,\dots,n\}$ with $r$ elements and over all sign reversals $\varepsilon = \{+1,-1\}$. The second sum runs over all strictly increasing sequences of subsets in $J$.

$J^C$ is a complement of $J$ with respect to $\{1,\dots,n\}$,
\begin{equation}
    \begin{aligned}
    h_{\varepsilon J;K} := \prod_{j\in J} \frac{1}{\eta}{(1-ax_j^{\varepsilon_j})(1-bx_j^{\varepsilon_j})(1-cx_j^{\varepsilon_j})(1-dx_j^{\varepsilon_j})\over (1-x_j^{2\varepsilon_j})(1-q^2 x_j^{2\varepsilon_j})} \times \prod_{\substack{j,j' \in J \\ j < j'}} \frac{1}{t^2} \frac{(1-t^2 x_j^{\varepsilon_j} x_{j'}^{\varepsilon_{j'}})}{(1- x_j^{\varepsilon_j} x_{j'}^{\varepsilon_{j'}})} \frac{(1- q^2 t^2 x_j^{\varepsilon_j} x_{j'}^{\varepsilon_{j'}})}{(1- q^2 x_j^{\varepsilon_j} x_{j'}^{\varepsilon_{j'}})} \times  \\
    \times  \prod_{\substack{j \in J \\ k \in K}} \frac{1}{t^2} \frac{(1-t^2 x_j^{\varepsilon_j} x_{k})}{(1- x_j^{\varepsilon_j} x_{k})} \frac{(1-t^2 x_j^{\varepsilon_j}/ x_{k})}{(1- x_j^{\varepsilon_j}/ x_{k})}
     \end{aligned}
\end{equation}

\section{Systems of type $C^\vee C$: eigenfunctions}

\subsection{Eigenfunctions: generality and examples}

The eigenfunctions have a triangular structure given as follows: for $n$ variables, one considers monomials $x^\alpha$ with $n$ integer $\alpha_i$'s (not obligatory positive). The level of $\alpha$ is defined $|\alpha|:=\sum_i|\alpha_i|$, and one can define partial orders on $\alpha$'s as follows. First of all, for a given set of integers $\alpha$ denote $\alpha^+$ a set $\{|\alpha_i|\}$ ordered to form a Young diagram, i.e. put in a weakly descending order. One can produce all $\alpha$ with the same $|\alpha|$ by acting on $\alpha^+$ by elements of the finite Weyl group $W$ generated by elements $s_i(\alpha)$, $i=1,\ldots,n$ such that
\be
s_i:(\alpha_1,\ldots,\alpha_{i-1},\alpha_i,\alpha_{i+1},\ldots,\alpha_n)&\to&
(\alpha_1,\ldots,\alpha_{i-1},\alpha_{i+1},\alpha_i,\ldots,\alpha_n),\ \ \ \ \ \ \ \ \ \  i=1,\ldots,n-1\nn\\
s_n:(\alpha_1,\ldots,\alpha_{n-1},\alpha_n)&\to& (\alpha_1,\ldots,\alpha_{n-1},-\alpha_n)
\ee
We will also need the affine Weyl group $\widehat W$ which differs from the finite Weyl group $W$ by adding the affine element $s_0$:
\be
s_0:(\alpha_1,\alpha_2,\ldots,\alpha_n)\to(-\alpha_1-1,\alpha_2,\ldots,\alpha_n)
\ee
Then, there are two ways to define partial orders:
\begin{itemize}
\item[1)] For any $\alpha$, define $w_\alpha\in \widehat W$ to be the shortest element such that $w_\alpha(\alpha)=0$. Then,
$\alpha\succ\beta$ if $w_\alpha$ is longer than $w_\beta$.
\item[2)] For $\alpha=w(\alpha^+)$, $\beta=w'(\beta^+)$, $w,w'\in W$, one defines
$\alpha>\beta$ if $\alpha^+>\beta^+$, or, when $\alpha^+=\beta^+$, if the minimal length of $w$ is less than that of $w'$.
\end{itemize}
Now the eigenfunctions of the Cherednik Hamiltonians $E^C_\alpha$ have triangular expansions with respect to the both orders:
\be\label{triang}
E^C_\alpha=x^\alpha+\sum_{\beta\prec\alpha}C^{(1)}_{\alpha\beta}x^\beta=x^\alpha+\sum_{\beta<\alpha}C^{(2)}_{\alpha\beta}x^\beta
\ee
Hereafter we normalize the polynomials to have unit coefficient in front of the leading term, they are called monic non-symmetric Koornwinder polynomials, and similarly monic symmetric Koornwinder polynomials made from them (see below).

The first partial order restricts the sum (\ref{triang}) stronger (i.e. contains less number of $\beta$), and $C^{(1)}_{\alpha\beta}$ are all non-zero because of the Sahi theorem \cite[Th.5.1]{Sahi2}: $C^{(1)}_{\alpha\beta}\ne 0$ iff $\beta\prec\alpha$. At the same time, among $C^{(2)}_{\alpha\beta}$ there may be zeroes (when $\alpha$ and $\beta$ can not be compared w.r.t. to the first order, but can, to the second one).

As an example, at $|\alpha|=1$ and $n=2$, the four $\alpha$'s are ordered as $[1,0]>[0,1]=s_1[1,0]>[0,-1]=s_2s_1[1,0]>[-1,0]=s_1s_2s_1[1,0]$. More subtle example is the pair $[-1,1]$ and $[1,0]$. In the first order, the elements $w_{[1,0]}=s_0s_1s_2s_1$ and $w_{[-1,1]}=s_0s_1s_2s_0$ have the same length, and, hence, the monomial $x_1$ is absent in $E^C_{[-1,1]}$, which is, indeed, the case. At the same time, $[-1,1]>[1,0]$ in the second order.

Note that, in the $A_n$ DAHA, there are two similar orders\footnote{Again, e.g. at $n=3$, elements $w_{[002]}=s_0^{-1}s_2s_1s_0^{-1}$ and $w_{[110]}=s_0^{-1}s_0^{-1}s_1s_2$ have the same length, and the monomial $x_1x_2$ is absent in $E^C_{[0,0,2]}$.
}. We do not go into any further details here, since we just define the non-symmetric Koornwinder polynomials as eigenfunctions of the Cherednik Hamiltonians,
\be\label{ES}
C_iE^C_\alpha=\Lambda_\alpha^{(i)}E^C_\alpha
\ee
and vanishing of some coefficients $C^{(2)}_{\alpha\beta}$ in (\ref{triang}) is automatically achieved. The only important point is that the leading coefficient $x^\alpha$ in $E^C_\alpha$ is always non-zero.

The first examples of the polynomials at $n=2$ are:
\begin{itemize}
\item The ground state:
\be
E^C_{[0,0]}=1
\ee
\item The first excitations are
\be
E^C_{[-1,0]}&=&{1\over x_1}-t^2{e_1(a,c)e_2(b,d)t^2-e_1(b,d)\over e_4(a,b,c,d)t^4-1}\nn\\
E^C_{[0,-1]}&=&{1\over x_2}+{e_4(a,b,c,d)(t^2-1)\over e_4(a,b,c,d)t^2-1}{1\over x_1}-{e_1(a,c)e_2(b,d)t^2-e_1(b,d)\over e_4(a,b,c,d)t^2-1}\nn\\
E^C_{[0,1]}&=&x_2+t^2{e_4(a,b,c,d)q^2t^2+e_2(a,c)(q^2t^2-1)-q^2\over e_4(a,b,c,d)q^2t^4-1}{1\over x_2}+
{(t^2-1)(e_2(b,d)q^2t^2-1)e_2(a,c)\over e_4(a,b,c,d)q^2t^4-1}{1\over x_1}-\nn\\
&-& t^2{e_3(a,b,c,d)q^2t^2-e_1(b,d)q^2-e_1(a,c)\over e_4(a,b,c,d)q^2t^4-1}\nn\\
E^C_{[1,0]}&=&x_1+q^2{t^2-1\over q^2t^2-1}\Big(x_2+{1\over x_2}\Big)+\nn\\
&+&{e_4(a,b,c,d)q^2t^2(q^2t^4-q^2t^2+q^2-1)+e_2(a,c)(q^2-1)(q^2t^4-1)+q^2(1-q^2t^2)\over(q^2t^2-1)
(e_4(a,b,c,d)q^2t^4-1)}{1\over x_1}-\nn\\
&-&{q^2t^4-1\over q^2t^2-1}{q^2t^2e_3(a,b,c,d)-q^2e_1(b,d)-e_1(a,c)\over e_4(a,b,c,d)q^2t^4-1}
\ee
where $e_k$ are the elementary symmetric polynomials.
\end{itemize}

One can construct symmetric combinations of the non-symmetric Koornwinder polynomials, and, similarly to the $A_n$ case, these combinations are constructed from polynomials corresponding to one and the same Young diagram.
In the example above, the symmetric polynomial $K_{[1]}$ is
\be\label{K1}
K_{[1]}=E^C_{[1,0]}+{q^2-1\over q^2t^2-1}E^C_{[0,1]}+(q^2-1){e_2(a,c)t^2-1\over e_4(a,b,c,d)q^2t^4-1}E^C_{[0,-1]}+\nn\\
+(q^2-1){e_2(a,c)t^2-1\over e_4(a,b,c,d)q^2t^4-1}{e_4(a,b,c,d)-1\over e_4(a,b,c,d)q^2t^4-1}E^C_{[-1,0]}=\nn\\
=x_1+x_2+{1\over x_1}+{1\over x_2}-(1+t^2){t^2e_3(a,b,c,d)-e_1(a,b,c,d)\over e_4(a,b,c,d)t^4-1}
\ee
Note how the coefficients in front of different $E_\alpha$ are all non-trivial and distinct. Nevertheless, as a result of a clever conspiracy, the resulting polynomial is symmetric. 

\subsection{Constructing non-symmetric polynomials recursively}

The eigenvalues associated with non-symmetric Koornwinder polynomials in equation (\ref{ES}) are (it is a counterpart of (\ref{evw}) in the $A_n$ case)
\be
\Lambda_\alpha^{(i)}=q^{2\alpha_i}\eta^{{\rm sgn}(\alpha_i)}t^{2\zeta_\alpha(i)}
\ee
where sgn$(0)=1$, and
\be
\zeta_\alpha(i):=[w(\rho)]_i
\ee
where $w\in W$ is the shortest element that produces $\alpha$ from $\alpha^+$: $\alpha=w(\alpha^+)$, and $\rho$ is the shifted Weyl vector with components $\rho_i=n-i$. It can be rewritten in the form
\begin{equation}
\zeta_{\alpha}(i) = \left\{
    \begin{aligned}
       & \# \left \{k<i: \begin{array}{l}
             |\alpha_i| \geq |\alpha_k|, \, \alpha_k<0\\
             |\alpha_i|> |\alpha_k|, \, \alpha_k>0
        \end{array} \right\}\left(1-\delta_{\alpha_i,0} \right) + \# \left\{k>i:\, |\alpha_i| \geq |\alpha_k| \right\} \quad &&\text{at} \quad \alpha_i\geq 0 \\
        - &\# \left \{k<i: \begin{array}{l}
             |\alpha_i| \geq |\alpha_k|, \, \alpha_k<0\\
             |\alpha_i|> |\alpha_k|, \, \alpha_k>0
        \end{array} \right\} - \# \left\{k>i:\, |\alpha_i|> |\alpha_k| \right\} \quad &&\text{at} \quad \alpha_i< 0
    \end{aligned} \right.
\end{equation}

Following \cite{CR}, one can obtain recursion relations for the polynomial eigenfunctions (in \cite{CR}, they are called the electronic Macdonald polynomials) similar to the Sahi-Knop recursion relations in the $A_n$ case. To this end, we introduce an operator (compare with $B$ in (\ref{B}))
\be
B:=t^{2(1-n)}T_1\ldots T_{n-1}T_nT_{n-1}\ldots T_1x_1
\ee
and the quantities ($i=1,\ldots,n-1$)
\be
r_{\alpha,0}:&=&{1\over\Lambda_\alpha^{(1)}}\nn\\
r_{\alpha,i}:&=&{\Lambda_\alpha^{(i+1)}\over\Lambda_\alpha^{(i)}}\nn\\
r_{\alpha,n}:&=&{1\over\Lambda_\alpha^{(n)}}
\ee
Then, the recursion relations are (cf. \cite[p.21]{CR}):
\begin{itemize}
\item Since $[C_1,B^{-1}]$ is the intertwining operator \cite{Sahi}, and
\be
[C_1,B^{-1}]E^C_\alpha\sim E^C_{s_0\alpha}
\ee
one has
\be\label{BEC}
\begin{array}{rcll}
BE^C_\alpha&=&q^2\displaystyle{\eta r_{\alpha,0}(a^{-1}+c^{-1})-b-d\over r_{\alpha,0}^2-q^2}E^C_\alpha+\cr
&+&
\displaystyle{bdq^{2\alpha_1}r_{\alpha,0}\over \eta}
\displaystyle{(r_{\alpha,0}-cd\eta^{-1})(r_{\alpha,0}-ad\eta^{-1})(r_{\alpha,0}-ab\eta^{-1})(r_{\alpha,0}-bc\eta^{-1})\over
(r_{\alpha,0}^2-q^2)^2}E^C_{s_0\alpha},&\ \ \hbox{if}\ \ \alpha_1<0\cr\cr
BE^C_\alpha&=&q^2\displaystyle{\eta r_{\alpha,0}(a^{-1}+c^{-1})-b-d\over r_{\alpha,0}^2-q^2}E^C_\alpha
-q^{2\alpha_1}\eta r_{\alpha,0}E^C_{s_0\alpha},&\ \ \hbox{if}\ \ \alpha_1\ge0
\end{array}
\ee
It is a counterpart of much simpler (\ref{BE}) in the $A_n$ case.
\item Similarly, at $i=1,\ldots,n-1$,
\be\label{TE}
\begin{array}{rcll}
T_iE^C_\alpha&=&t^2E^C_\alpha,&\ \ \ \ \ \ \hbox{if}\ \ \ \ \ \alpha_i=\alpha_{i+1}\cr\cr
T_iE^C_\alpha&=&-\displaystyle{t^2-1\over r_{\alpha,i}-1}E^C_\alpha+t^2E^C_{s_i\alpha},&
\ \ \ \ \ \ \hbox{if}\ \ \ \ \ \alpha_i<\alpha_{i+1}\cr\cr
T_iE^C_\alpha&=&-\displaystyle{t^2-1\over r_{\alpha,i}-1}E^C_\alpha+
{(1-t^2r_{\alpha,i})(1-t^{-2}r_{\alpha,i})\over (1-r_{\alpha,i})^2}E^C_{s_i\alpha},&\ \ \ \ \ \ \hbox{if}\ \ \ \ \ \alpha_i>\alpha_{i+1}
\end{array}
\ee
It is the same as in the non-twisted $A_n$ case of (\ref{perm})-(\ref{C12}), up to a distinct normalization of $T_i$'s.
\item At last,
\be\label{TnE}
\begin{array}{rcll}
T_nE^C_\alpha&=&bdE^C_\alpha,&\ \ \ \ \ \ \hbox{if}\ \ \ \ \ \alpha_n=0\cr\cr
T_nE^C_\alpha&=&\displaystyle{r_{\alpha,n}(\eta+bd\eta^{-1})-(bd+1)\over r_{\alpha,n}^2-1}E^C_\alpha+bdE^C_{s_n\alpha},&
\ \ \ \ \ \ \hbox{if}\ \ \ \ \ \alpha_n<0\cr\cr
T_nE^C_\alpha&=&\displaystyle{r_{\alpha,n}(\eta+bd\eta^{-1})-(bd+1)\over r_{\alpha,n}^2-1}E^C_\alpha-&\cr
&-&\displaystyle{(\eta^{-1}r_{\alpha,n}-1)(bd\eta^{-1}r_{\alpha,n}-1)(acq^{-2}\eta^{-1}r_{\alpha,n}-1)(\eta r_{\alpha,n}-1)
\over (r_{\alpha,n}^2-1)^2}E^C_{s_n\alpha},&\ \ \ \ \ \ \hbox{if}\ \ \ \ \ \alpha_n>0
\end{array}
\ee
\end{itemize}

This set of relations allows one to construct all polynomials recursively using (\ref{BEC}) at $\alpha_1>0$, (\ref{TE}) at $\alpha_i<\alpha_{i+1}$ ($i=1,\ldots,n-1$) and (\ref{TnE}) at $\alpha_n<0$. For instance, all $\alpha$ up to level 2 can be generated as (though even here the minimal possible path to a given $\alpha$ is sometimes not unique):
\be
[0,0]\stackrel{B}{\longrightarrow}[-1,0]\stackrel{T_1}{\longrightarrow}[0,-1]\stackrel{T_2}{\longrightarrow}&[0,1]&
\stackrel{T_1}{\longrightarrow}[1,0]\stackrel{B}{\longrightarrow}[-2,0]\stackrel{T_1}{\longrightarrow}[0,-2]
\stackrel{B}{\longrightarrow}[0,2]\stackrel{T_1}{\longrightarrow}[2,0]\nn\\
&\downarrow_B&\nn\\
&[-1,1]&\stackrel{T_1}{\longrightarrow}[1,-1]\stackrel{T_2}{\longrightarrow}[1,1]\nn\\
&\downarrow_{T_2}&\nn\\
&[-1,-1]&\nn
\ee

\subsection{Symmetric Koornwinder polynomials\label{symmsec}}

Similarly to the $A_n$ case, one can produce symmetric polynomials by summing the non-symmetric ones over the same orbit $\alpha^+$
(cf. \cite[Prop.4.4]{CR}):
\be\label{symm}
K_{\alpha^+}=\sum_{{\beta=w(\alpha^+)}\atop{w\in W}}c(\alpha^+,\beta)E^C_\beta
\ee
and we normalize the polynomial so that the coefficient in front of $x^{\alpha^+}$ is unity:
\be\label{Ktr}
K_{\lambda}=x^{\lambda}+\sum_{\mu<\lambda}c_{\lambda\mu}x^\mu
\ee
where the sum runs over lexicographically ordered Young diagrams.

Formula (\ref{symm}) is a counterpart of the $A_n$ case formula (\ref{nss}), however, the coefficients $c(\alpha^+,\beta)$ are a bit more involved, and can be again evaluated by the following procedure: first, one obtains $\beta$ from $\alpha^+$ by a minimal length signed permutation, $\sigma\in W$, which is a product of transpositions $\sigma=\prod_k s_{i_k}$, $i_k\in [1,\ldots,n]$. Then, one associates with each transposition $s_i$ acting on $\alpha$ a multiplier $N_\alpha(s_i)$:
\begin{itemize}
\item For transpositions $s_i:(\alpha_1,\ldots,\alpha_{i-1},\alpha_i,\alpha_{i+1},\ldots,\alpha_n)\to
(\alpha_1,\ldots,\alpha_{i-1},\alpha_{i+1},\alpha_i,\ldots,\alpha_n)$, $i=1,\ldots,n-1$, $\alpha_i>\alpha_{i+1}$, the multiplier is
\be
N_\alpha(s_i)={\Lambda^{(i+1)}_\alpha-t^{-2}\Lambda^{(i)}_\alpha\over
\Lambda^{(i+1)}_\alpha-\Lambda^{(i)}_\alpha}={1\over t^2}{1-t^2r_{\alpha,i}\over 1-r_{\alpha,i}}
\ee
This formula is a literal counterpart of (\ref{N}) in the $A_n$ case.
\item For the transposition $s_n:(\alpha_1,\ldots,\alpha_{n-1},\alpha_n)\to (\alpha_1,\ldots,\alpha_{n-1},-\alpha_n)$, it is
\be
N_\alpha(s_n)={1\over\eta}{\left(r_{\alpha,n}-\displaystyle{\eta\over bd}\right)
\Big(1-\eta r_{\alpha,n}\Big)\over 1-r_{\alpha,n}^2}
\ee
This is a new element as compared with the $A_n$ case corresponding to the new element in the Weyl group of the $BC_n$ series. Note that $\Lambda^{(i)}_\alpha\Lambda^{(i)}_{s\alpha}=1$.
\end{itemize}
Now the coefficients in (\ref{symm}) are
\be
c(\alpha^+,\beta)=\prod_k N_{\alpha^{(k)}}(s_{i_k})
\ee
where $\alpha^{(k)}$ is the partition obtained after application of $k-1$ transpositions so that
  $\alpha^{(1)} = \alpha^+$, $\alpha^{(2)} = s_{i_1} \alpha^+$, and so on.

\subsection{Orthogonality}

There is a scalar product on the symmetric Koornwinder polynomials \cite{Koorn},
\be
\Big<f\Big|g\Big>_s&=&{1\over n!}\oint_0\prod_{i=1}^n{dx_i\over x_i}f(x)g(x^{-1})\Delta_s(x)\Delta_s(x^{-1}),\nn\\
\Delta_s(x):&=&\prod_{i=1}^n{(x_i^2;q^2)_\infty\over (ax_i,bx_i,cx_i,dx_i;q^2)_\infty}
\prod_{i<j}{(x_ix_j;q^2)_\infty\over(t^2x_ix_j;q^2)_\infty}{(x_ix_j^{-1};q^2)_\infty\over(t^2x_ix_j^{-1};q^2)_\infty}
\ee
so that the polynomials are orthogonal w.r.t. this scalar product:
\be
\Big<K_\lambda\Big|K_\mu\Big>_s\sim\delta_{\lambda\mu}
\ee
This allows one to unambiguously obtain them using the triangular expansion (\ref{Ktr}).

A similar scalar product also exists for the non-symmetric Koornwinder polynomials \cite[Sec.3]{Sahi2}:
\be\label{nssp}
\Big<f\Big|g\Big>_{ns}&=&{1\over n!}\oint_0\prod_{i=1}^n{dx_i\over x_i}f(x)\tilde{g}(x^{-1})\Delta_{ns}(x),\nn\\
\Delta_{ns}(x):&=&\Delta_s(x)\Delta_s(x^{-1})
\prod_{i=1}^n(q^2x_i^2;q^2)_\infty(q^2x_i^{-2};q^2)_\infty\prod_{i=1}^n{(x_i-a)(x_i-c)\over x_i^2-1}
\prod_{i<j}{(x_ix_j-t^2)(x_ix_j^{-1}-t^2)\over (x_ix_j-1)(x_ix_j^{-1}-1)}\nn\\
\Big<E^C_\alpha\Big|E^C_\beta\Big>_{ns}&\sim&\delta_{\alpha\beta}
\ee
where tilde denotes that one has to substitute $(a,b,c,d,q,t)\to(a^{-1},b^{-1},c^{-1},d^{-1},q^{-1},t^{-1})$.
However, it is technically less effective in evaluating the polynomials (though the triangular expansion (\ref{triang}) exists in this case as well), since it involves the polynomials given both at $a,b,c,d,q,t$ and at their inverse.

Equivalently, one can use the pair $b,d$ instead of $a,c$ in the definition of $\Delta_{ns}(x)$.

\subsection{Evaluation}

The non-symmetric Koornwinder polynomials are factorized at the point $1/\overline{\Lambda}^{(i)}_\emptyset$.

The evaluation formula in the dual terms looks like \cite[Th.9.3]{St}:
\be\label{Ef}
E^C_\alpha\left({1\over\overline{\Lambda}^{(i)}_\emptyset}\right)=F_{\alpha^+}^{(0)}\prod_k F_\alpha(s_{i_k})
\ee
(see the definition of bars in (\ref{dualAW})).
Here again, similarly to sec.\ref{symmsec}, one has to obtain $\alpha$ from $\alpha^+$ by a minimal length signed permutation, $\sigma\in W$, which is a product of transpositions $\sigma=\prod_k s_{i_k}$, $i_k\in [1,\ldots,n]$, and then, one associates with each transposition a multiplier $F_\alpha(s_i)$ (note the striking similarity with $N_\alpha(s_i)$):
\begin{itemize}
\item For transpositions $s_i:(\alpha_1,\ldots,\alpha_{i-1},\alpha_i,\alpha_{i+1},\ldots,\alpha_n)\to
(\alpha_1,\ldots,\alpha_{i-1},\alpha_{i+1},\alpha_i,\ldots,\alpha_n)$, $i=1,\ldots,n-1$, $\alpha_i>\alpha_{i+1}$, the multiplier is
\be
F_\alpha(s_i)={1\over t^2}{1-t^{-2}r_{\alpha,i}\over 1-r_{\alpha,i}}
\ee
\item For the transposition $s_n:(\alpha_1,\ldots,\alpha_{n-1},\alpha_n)\to (\alpha_1,\ldots,\alpha_{n-1},-\alpha_n)$, it is
\be
F_\alpha(s_n)={1\over\eta}{\left(r_{\alpha,n}-\displaystyle{\eta\over bd}\right)
\Big(r_{\alpha,n}-\eta\Big)\over 1-r_{\alpha,n}^2}
\ee
\item At last,
\be
F_{\lambda}^{(0)}=\prod_{i=1}^n{(adt^{2(n-i)},q^2bdt^{2(n-i)},cdt^{2(n-i)},
abcdt^{2(n-i)};q^2)_{\lambda_i}\over d^{\lambda_i}t^{2(n-i)\lambda_i}(abcdt^{4(n-i)};q^2)_{2\lambda_i}}\times\nn\\
\times\prod_{1\le i< j\le n}{(q^2t^{2(j-i+1)};q^2)_{\lambda_i-\lambda_j}\over(q^2t^{2(j-i)};q^2)_{\lambda_i-\lambda_j}}
{(abcdt^{2(2n-i-j+1)};q^2)_{\lambda_i+\lambda_j}\over(abcdt^{2(2n-i-j)};q^2)_{\lambda_i+\lambda_j}}
\ee
where we use the standard notation for the $q$-Pochhammer symbol:
\be
(x;q)_n:=\prod_{j=0}^{n-1}(1-xq^j),\ \ \ \ \ \ \ \ \ (x_1,x_2,\ldots,x_k;q)_n:=\prod_{i=1}^k(x_i;q)_n
\ee
$F_{\alpha^+}$ is a new element as compared with sec.\ref{symmsec}, where the coefficient in front of term associated with the Young diagram is 1, and here it is $F_{\alpha^+}$.
\end{itemize}

The symmetric Koornwinder polynomials are also factorized at the same point, and also at the point $\overline{\Lambda}^{(i)}_\emptyset$ because of the symmetry properties \cite[Remark 9.5]{St}:
\be
K_\lambda\left(\overline{\Lambda}^{(i)}_\emptyset\right)=\prod_{i=1}^n{(adt^{2(n-i)},bdt^{2(n-i)},cdt^{2(n-i)},
abcdq^{-2}t^{2(n-i)};q^2)_{\lambda_i}\over d^{\lambda_i}t^{2(n-i)\lambda_i}(abcdq^{-2}t^{4(n-i)};q^2)_{2\lambda_i}}\times\nn\\
\times\prod_{1\le i< j\le n}{(t^{2(j-i+1)};q^2)_{\lambda_i-\lambda_j}\over(t^{2(j-i)};q^2)_{\lambda_i-\lambda_j}}
{(abcdq^{-2}t^{2(2n-i-j+1)};q^2)_{\lambda_i+\lambda_j}\over(abcdq^{-2}t^{2(2n-i-j)};q^2)_{\lambda_i+\lambda_j}}
\ee
This formula is much similar to $F_{\alpha^+}$ differing only by additional factors of $q^{-2}$ in some of the $q$-Pochhammer symbols.

\subsection{Duality relations}

There is a duality, which is a counterpart of the standard duality of the Macdonald polynomials of type $A$, see sec.\ref{sdua}.

Its counterparts in terms of the non-symmetric Koornwinder polynomials is the involution transform $T_0\to B^{-1}$, $T_i\to T_i$, $x_i\leftrightarrow C_i$ ($i=1,\ldots,n$) \cite{Sahi}, and is of the form
\be \label{EE}
{E^C_\alpha\left({1\over\overline{\Lambda}^{(i)}_\beta}\right)\over
E^C_\alpha\left({1\over\overline{\Lambda}^{(i)}_\emptyset}\right)}={\overline{E}^C_\beta\left({1\over\Lambda^{(i)}_\alpha}\right)\over
\overline{E}^C_\beta\left({1\over\Lambda^{(i)}_\emptyset}\right)}
\ee
while, for the symmetric Koornwinder polynomials, it looks like \cite{Sahi} (see also \cite{CherednikConj,CherednikDAHA})
\be\label{Kdu}
{K_{\alpha^+}\left(\overline{\Lambda}^{(i)}_{\beta^+}\right)\over
K_{\alpha^+}\left(\overline{\Lambda}^{(i)}_{\emptyset}\right)}={\overline{K}_{\beta^+}\left(\Lambda^{(i)}_{\alpha^+}\right)\over
\overline{K}_{\beta^+}\left(\Lambda^{(i)}_{\emptyset}\right)}
\ee
Because of the symmetry of polynomials $K$ one can write this equality at inverse eigenvalues similarly to (\ref{EE}).

Note that these are exactly the quantities entering the r.h.s. of the CMM formulas \cite{CheCMM}.

Note also that the duality implies that the quantities $\overline{E}^C_\alpha\left({1\over\Lambda^{(i)}_\emptyset}\right)$ are factorized, which is generated from (\ref{Ef}) by the duality.

\subsection{Symmetric polynomials: manifest expressions\label{me}}

Now we give a few explicit examples of the symmetric Koornwinder polynomials.
Introduce the symbol
\be\label{111}
\langle x;a\rangle:=\{xa\}\{xa^{-1}\}
\ee
and the symmetric $q$-Pochhammer symbols:
\be\label{112}
 \{x;q\}_k:=\prod_{j=0}^{k-1}\{q^jx\},\ \ \ \ \ \ \ \ \ \ \{x_1,\ldots,x_l;q\}_k:=\prod_{i=1}^l\{x_i;q\}_k\ \ \ \ \ \ \ \ \ \
 \langle x;a\rangle_k:=\prod_{j=0}^{k-1}\langle x;q^ja\rangle
\ee
We also define $W$-invariant Laurent polynomials
\be
E_k(x_1,\ldots,x_n):=\sum_{1\le j_1<\ldots<j_k\le n}\prod_{r=1}^k\langle\sqrt{x};t^{j_r-r}\sqrt{a}\rangle
\ee
and
\be
H_k(x_1,\ldots,x_n):=\sum_{|\nu|=k}\prod_{i=1}^n{\{t;q\}_{\nu_i}\over\{q;q\}_{\nu_i}}
\langle\sqrt{x_i};\sqrt{a}t^{i-1}q^{\sum_{j=2}^i\nu_{j-1}}\rangle_{\nu_i}
\ee
which are counterparts of the elementary and complete homogeneous symmetric polynomials accordingly.

There is a counterpart of the Cauchy formula \cite{Mimachi}, which reads, for any positive $m$ and $n$,
\be
\sum_{\lambda\subset [n^m]}(-1)^{nm+|\lambda|}K_{\lambda}(x_1,\ldots,x_m;q,t)K_{[m^n/\lambda^\vee]}(y_1,\ldots,y_n;t,q)=
\prod_{i=1}^m\prod_{j=1}^n\langle\sqrt{x_i};\sqrt{y_j}\rangle
=\prod_{i=1}^m\prod_{j=1}^n\Big(x_i+{1\over x_i}-y_j-{1\over y_j}\Big)\nn\\
\ee
where $[m^n/\mu]$ denotes the Young diagram $[m-\mu_n,\ldots,m-\mu_1]$.

Then,
\be
K_{[1^r]}(x_1,\ldots,x_n)=\sum_{j=0}^r{\{t^{n-r+1},\sqrt{ab}t^{n-r},\sqrt{ac}t^{n-r},\sqrt{ad}t^{n-r};t\}_j\over
\{t,\eta qt^{2(n-r)};t\}_l}E_{r-l}(x_1,\ldots,x_n)
\ee
and, similarly,
\be
K_{[r]}(x_1,\ldots,x_n)={\{t^n,\sqrt{ab}t^{n-1},\sqrt{ac}t^{n-1},\sqrt{ad}t^{n-1};q\}_r\over\{t,\eta q^rt^{2(n-1)};q\}_r}\times\nn\\
\times\sum_{j=0}^r(-1)^j{\{q^{-r},\eta q^rt^{2(n-1)};q\}_j\over\{t^n,\sqrt{ab}t^{n-1},\sqrt{ac}t^{n-1},\sqrt{ad}t^{n-1};q\}_j}
H_j(x_1,\ldots,x_n)
\ee
For instance,
\be
K_{[1]}(x_1,x_2)&\stackrel{(\ref{K1})}{=}&E_1(x_1,x_2)+{(1+t^2)(abt^2-1)(act^2-1)(adt^2-1)\over at^2(abcdt^4-1)}\nn\\
K_{[1]}(x_1,\ldots,x_n)&=&E_1(x_1,\ldots,x_n)+{(1-t^{2n})(abt^{2(n-1)}-1)(act^{2(n-1)}-1)(adt^{2(n-1)}-1)\over at^{2(n-r)}(1-t^2)(abcdt^{4(n-1)}-1)}
\ee

\subsection{Stability}

The non-symmetric Koornwinder polynomials celebrate a weak form of the stability property:
\be
\Bigg[E^C_{[\alpha_1,\ldots,\alpha_{n-1},0]}\Bigg]_{x_n,0}=E^C_{[\alpha_1,\ldots,\alpha_{n-1}]}\ \ \ \ \ \ \ \ \ \ \hbox{at } t=1
\ee
where $[...]_{x,k}$ denotes the $k$-th power of $x$.

This property translates to the symmetric Koornwinder polynomials in the form:
\be
\Bigg[K_{[\alpha_1,\ldots,\alpha_{n-1},0]}\Bigg]_{x_n,0}\approx K_{[\alpha_1,\ldots,\alpha_{n-1}]}
\ \ \ \ \ \ \ \ \ \ \hbox{at } t=1
\ee
where $\alpha$ is now a Young diagram, and $\approx$ means that the two Laurent polynomials coincide up to numerical factors (independent of $a$, $b$, $c$, $d$ and $q$) in some coefficients. To be more precise, for partitions $\lambda$ and $\mu$ of lengths $n-1$, if
\be
\Bigg[K_{[\lambda,0]}\Bigg]_{x_n,0}=\sum_\mu {\cal K}_{\lambda\mu}x^\mu
\ \ \ \ \ \ \ \ \ \ \hbox{at } t=1\nn
\ee
then
\be
K_{\lambda}=\sum_\mu \displaystyle{n-l_\lambda\over n-l_\mu}\ {\cal K}_{\lambda\mu}x^\mu\ \ \ \ \ \ \ \ \ \ \hbox{at } t=1
\ee
Examples of this property are clearly seen in the manifest expressions of sec.\ref{me}.

\section{Towards triad of type $C^\vee C$\label{5}}

The triad consists of three basic elements: of the universal solution, which is the NS power series in the $A$-series case, and of two polynomial reductions: those to symmetric polynomial, and to BA functions. In this section, we describe possible approaches to all of these elements, and we start with the BA function.

\subsection{Baker-Akhiezer (BA) function}

\subsubsection{BA function for reduced finite root system\label{frs}}

The non-symmetric polynomials are associated with DAHAs, which are, in turn, associated with affine root systems. At the same time, symmetric polynomials are rather associated with the spherical DAHA (or DIM-like structures), which means they are associated rather with finite root systems. As a particular manifestation of this, O.\,Chalykh \cite{Cha} proposed\footnote{In contrast with the original Chalykh's paper, we interchanged space and cospace in the definitions of (quasi)polynomial (\ref{BAgen}) and periodicity condition (\ref{BAgeneq}) so that the resulting BA functions would match the corresponding Macdonald polynomials.} a construction of the BA function suitable for any finite root system. To describe his construction, let us consider a Euclidean space $\mathbb{R}^n$ with the scalar product of vectors $\vec x$ and $\vec y$ denoted as $(x|y)$, and consider an arbitrary root system $\mathbb{R}^n \supset R = R_{+}\cup (-R_{+})$ with roots\footnote{The standard notation $\alpha$ for roots used in this subsection should not be confused with the notation $\alpha$ for labels of non-symmetric polynomials used throughout the text.} $\vec\alpha$, $R_{+}$ being positive roots, $\vec\alpha^{\vee} = 2\vec\alpha/(\alpha,\alpha)$ being coroots. Root system $R$ is symmetric under the action of Weyl group $W_R$, $\{m_{\alpha}\}$ denotes a set of parameters $m_{\alpha}$ for each orbit of the Weyl group. It means that roots with length $(\alpha,\alpha) = 2$ are associated with a parameter $m$, but if the root system also contains longer or shorter roots, they are assigned additional parameters $m_{\alpha}$. We also introduce a refined Weyl vector:
\begin{equation}\label{rW}
    \vec\rho_m = \frac{1}{2} \sum_{\alpha\in R_{+}} m_{\alpha} \vec\alpha
\end{equation}

Consider the case of all $m_\alpha$ being non-negative integers.
Then, the multivariable Baker-Akhiezer function $\Psi_{\{m_{\alpha}\}}(\vec{x},\vec{y};\,q)$ of $n$ variables $\vec x$ and $\vec y$: $x_i=q^{z_i}$, $y_i=q^{\lambda_i}$ is defined to be the (quasi)polynomial
\begin{equation}
    \Psi_{\{m_{\alpha}\}}(\vec{x},\vec{y};\,q) = q^{\left(\lambda +\rho_{m}|\,z\right)} \sum_{\{k_{\alpha}\}} \psi_{\{k_{\alpha}\}} q^{-\sum_{\alpha \in R_{+}}k_{\alpha} (\alpha|\,z) } \label{BAgen}
\end{equation}
that satisfies the periodicity conditions:
\begin{equation}
    \Psi_{\{m_{\alpha}\}}\Big(\vec xq^{{1\over 2} j\vec\alpha^{\vee}},\,\vec{y};\,q\Big) =    \Psi_{\{m_{\alpha}\}}\Big(\vec{x}q^{-{1\over 2} j\vec\alpha^{\vee}},\,\vec{y};\,q\Big) \quad \text{at} \quad q^{(\alpha|z)} = 1
    \label{BAgeneq}
\end{equation}
for $j = 1, \dots, m_{\alpha}$ and for each $\vec\alpha\in R$. The sum in (\ref{BAgen}) goes over a set $\{k_{\alpha}\}$ such that, for each positive root $\vec\alpha\in R_+$, $k_{\alpha} = 0,\dots,m_{\alpha}$.

Then, this BA function is proved \cite{Cha} to be
\begin{itemize}
\item uniquely determined up to a common normalization factor (which may depend on $y$);
\item an eigenfunction of a proper Macdonald Hamiltonian, see \cite[Eqs.[2.2),(3.7)]{Cha};
\item symmetric under the permutation of sets $\{x_i\}$ and $\{y_i\}$ if being normalized by the condition
\be\label{Norm}
\psi_{0}=\prod_{\vec\alpha\in R_+}\prod_{j=1}^{m_\alpha}\{q^{(\alpha|z)-j}\}
\ee
which is nothing but a BA version of the duality relation for the symmetric Macdonald polynomials (\ref{Mdua}).
\end{itemize}

One can generate the corresponding symmetric Macdonald polynomial by summation over the Weyl group $W_R$:
\be
M_{\lambda+\rho_m}(x;q,q^{-m_\alpha})=\sum_{w\in W_R}\Psi_{\{m_{\alpha}\}}(w\vec{x},\vec{y};\,q)
\ee
where the normalization of the BA function is chosen to be $\psi_0=1$, and ,hereafter, the Young diagram $\lambda+\rho_m$ denotes $[\lambda_1+(\vec\rho_m)_1,\ldots,\lambda_n+(\vec\rho_m)_n]$.
These Macdonald polynomials are just the Koornwinder symmetric polynomials at particular values of parameters $a,b,c,d$.

\subsubsection{Koornwinder BA function: non-reduced finite root system\label{KBA}}

O.\,Chalykh \cite[sec.6]{Cha} also proposed a similar description of the BA function for the non-reduced finite root system $BC_n$, which is an eigenfunction of the Koornwinder Hamiltonian (\ref{KH}). Hence, we call this function Koornwinder BA function. In this case, there are five parameters $(t,a,b,c,d)$, and one chooses them to be $t=q^{-m}$, $a=-q^{-m_1}$, $b=-q^{-m_2}$, $c=q^{-m_3}$, $d=q^{-m_4}$ with $m,m_i\in\mathbb{Z}_{\ge 0}$ and both $m_1+m_2=2M_1-1$ and $m_3+m_4=2M_2-1$ odd. In fact, the whole construction is symmetric in parameters $a,b,c,d$, the symmetry is broken by choosing two negative parameters, and the symmetry inside the pairs $(m_1,m_2)$ and $(m_3,m_3)$ preserves.

It turns out that the roots of the root system $B_n$ are sufficient for the definition of this BA function (see \cite{Koorn,Cha}). This root system consists of a set $R^l$ of long roots $\pm \varepsilon_i\pm \varepsilon_j$ ($1\leq i <j \leq n$), and of a set $R^s$ of short roots $\pm \varepsilon_i$ ($1\leq i \leq n$), $R=R^s+R^l$.
Then, we define the BA function to be a (quasi)polynomial
\begin{equation}\label{BAK}
    \Psi_{\{m_{\alpha}\} } (\vec x, \vec y;\, q) = q^{(\lambda+\rho_m|\,z)} \sum_{\{k_{\alpha}\}} \psi_{\{k_{\alpha}\}} q^{-\sum_{\vec\alpha\in R_{+}} k_{\alpha} (\alpha|z)},
\end{equation}
where $k_{\alpha} = 0,\dots m_{\alpha}$ for each positive root $\alpha\in R_{B_n}$, and we associate $m_\alpha=m$ with the long roots, and $m_\alpha=M_1+M_2$ with the short roots so that the refined Weyl vector is $(\vec\rho_m)_i=m(n-i)+{M_1+M_2\over 2}$.
 More explicitly:
\begin{equation}
    \Psi_{\{m_{\alpha}\} }(\vec x,\vec y ; \,q) = \prod_{i=1}^n x_i^{{\lambda_i\over 2}+m(n-i)+{M_1+M_2\over 2}} \sum_{\{k_i\}=0}^{m}\sum_{\{k_{ij},k'_{ij}\}=0}^{M_1+M_2}\psi_{\{k_i,k_{ij},k'_{ij}\}}
     \left( \prod_{i=1}^n\left(\frac{1}{x_i}\right)^{k_i}\prod_{1\leq i<j\leq n} \left(\frac{x_j}{x_i} \right)^{k_{ij}}\left(\frac{1}{x_i x_j} \right)^{k'_{ij}}  \right) \end{equation}
The periodicity conditions take the form:
\begin{align}\label{Kpc}
    & \Psi_{\{m_{\alpha}\} } (x_i q^{s}, \vec y;\, q) =   \Psi_{\{m_{\alpha}\} } (x_i q^{-s}, \vec y;\, q), \quad 0 < s \preccurlyeq (m_1,m_2) \quad \quad &&\text{at} \quad x_i = 1, \\
    & \Psi_{\{m_{\alpha}\} } (x_i q^{s}, \vec y;\, q) =   \Psi_{\{m_{\alpha}\} } (x_i q^{-s}, \vec y;\, q), \quad 0 < s \preccurlyeq (m_3,m_4) \quad \quad &&\text{at} \quad x_i =- 1, \\
    & \Psi_{\{m_{\alpha}\} } (x_i q^{s}, \vec y;\, q) =   \Psi_{\{m_{\alpha}\} } (x_j q^{s}, \vec y;\, q), \quad s = 1, \dots, k \quad \quad &&\text{at} \quad x_i = x_j, \\
    & \Psi_{\{m_{\alpha}\} } (x_i q^{s}, \vec y;\, q) =   \Psi_{\{m_{\alpha}\} } (x_j q^{-s}, \vec y;\, q), \quad s = 1, \dots, k \quad \quad &&\text{at} \quad x_i = x_j^{-1},
\end{align}
where $s \preccurlyeq (m_1,m_2)$ means that $s$ is not larger than the number of the same parity of $m_1$ and $m_2$ (of these two numbers, one is odd, and the other one is even obligatory).

Such defined BA function is again proved \cite{Cha} to be uniquely determined up to a common normalization factor, and is an eigenfunction of the Koornwinder Hamiltonian. Let us normalize the BA function by the condition
\be\label{NormK}
\psi_0=\prod_{\vec\alpha\in R_+^s}\ \prod_{0<j\preccurlyeq(m_1,m_2)}\{q^{(\alpha|z)-j}\}
\prod_{\vec\alpha\in R_+^s}\ \prod_{0<j\preccurlyeq(m_3,m_4)}\{q^{(\alpha|z)-j}\}
\prod_{\vec\alpha\in R_+^l}\ \prod_{j=1}^{m}\{q^{(\alpha|z)-j}\}
\ee

Such normalized BA function satisfies the symmetricity condition, which this time requires coming to the dual AW variables:
\be\label{Ksym}
\Psi_{\{m_{\alpha}\}}(\vec{x},\vec{y};\,q)=\overline{\Psi}_{\{m_{\alpha}\}}(\vec{y},\vec{x};\,q)
\ee
Note that this is nothing but a BA version of the duality relation for the symmetric Koornwinder polynomials (\ref{Kdu}).

One can also generate the symmetric Koornwinder polynomial by summation over the Weyl group $W_{B_n}$:
\be
K_{\lambda+\rho_m}(x;q,q^{-m_\alpha})=\sum_{w\in W_{B_n}}\Psi_{\{m_{\alpha}\}}(\vec{wx},\vec{y};\,q)
\ee
where the BA function is taken at the normalization $\psi_0=1$.

Note that, though, when constructing this BA function, we used the root system $B_n$, the very construction is different from that of the $B_n$ BA function (sec.\ref{frs}), and the Koornwinder BA function differs from the $B_n$ BA function (they coincide only upon a peculiar choosing of parameters, see sec.\ref{BBA} below).

\subsection{Symmetric polynomials: branching rules\label{secbr}}

In the case of type $A$ root systems, the branching rule (\ref{br}) has two essential properties: the coefficient (\ref{brc}) factorizes, and, which is more important, its structure automatically guarantees that only the Young diagrams $\nu\subset\mu$ contribute, and one does not need to choose the region of summation ``by hands''. The branching rule for the type $C^\vee C$ is also known \cite{vDE}, however, the both properties are absent. The branching rule is also much more involved, which is not that surprising since the symmetric polynomials are no longer graded in the $C^\vee C$ case. A natural combination that substitutes now degrees of $x_1$ in the expansion (\ref{br}) is $\langle x_1;t_0\rangle_k$, and the branching rule is
\be\label{brK}
K_\mu(x_1,\ldots,x_n;q,t,a,b,c,d)=\sum_{\nu\curlyeqprec\mu} K_{\mu/\nu}(x_1;q,t,a,b,c,d)K_\nu(x_2,\ldots,x_{n};q,t,a,b,c,d)
\ee
where the relation $\nu\curlyeqprec\mu$ means that there is a Young diagram $\tau$ such that $\nu\subset\tau\subset\mu$ and $\mu/\tau$, $\nu/\tau$ are horizontal strips, i.e. $\mu$ can be obtained from $\nu$ by adding at most two horizontal strips. Here the skew polynomial $K_{\mu/\nu}(x_1;q,t,a,b,c,d)$ is given by the formula
\be
K_{\mu/\nu}(x_1;q,t,a,b,c,d)=\sum_{k=0}^{\#i:\mu^\vee_i=\nu^\vee_i+1}(-1)^{k+|\mu|+|\nu|}
C^{[(n+1)^m]-\mu^\vee,m}_{[n^m]-\nu^\vee,m-k}(q,t,a,b,c,d)\langle x_1;t_0\rangle_k
\ee
where $m:=\mu_1^\vee$, and  $C^{\mu,n}_{\lambda,r}$ are the coefficients in the counterpart of the Pieri formula for the Koornwinder polynomials,
\be
E_r(x_1,\ldots,x_n;q,t,a,b,c,d)\cdot K_\mu(x_1,\ldots,x_n;q,t,a,b,c,d)=\sum_\nu C^{\nu,n}_{\mu,r}K_\nu(x_1,\ldots,x_n;q,t,a,b,c,d)
\ee
They are manifestly given by a very involved formula, see \cite[Eq.(2.4)]{vDE}.

One can definitely repeat the way of obtaining formula (\ref{MNS}) for the Koornwinder polynomial, and obtain
\be\label{KNS}
K_\mu(x_1,\ldots,x_n;q,t,a,b,c,d)=\sum_{\nu_n\curlyeqprec\ldots\nu_1\curlyeqprec\nu_0:=\mu}\prod_{i=1}^n
K_{\nu_{i-1}/\nu_i}(x_i;q,t,a,b,c,d)
\ee
However, the summation domain is here restricted ``by hands'' and, hence, one can not just make $\mu_i$'s arbitrary complex numbers. In fact, there is an additional problem with this extension: to make $\mu_i$'s non-integer, one has to add some more terms so that the Koornwinder function becomes an eigenfunction of the Koornwinder Hamiltonian.

Since these formulas are very complicated, in the next subsection we demonstrate these problems with continuing them to a counterpart of the NS functions with arbitrary complex $\mu_i$'s.

\subsection{$C^\vee C_1$: Askey-Wilson (AW) triad\label{5.3}}

\subsubsection{AW polynomials}

The Koornwinder polynomial of one variable does not depend on the parameter $t$, and it is nothing but the AW polynomial \cite{AW}, manifestly given by the formula
\be\label{AWP}
K_{[n]}(x)={(ab,ac,ad;q^2)_n\over a^n(q^{2(n-1)}abcd;q^2)_n}\cdot \sum_{k=0}^n{(q^{-2n},q^{2(n-1)}abcd,ax,ax^{-1};q^2)_k\over (q^2,ab,ac,ad;q^2)_k}q^{2k}
\ee
where $n$ is integer. An essential feature of this polynomial as compared with (\ref{M2}) is that the finite sum goes over powers of $q$, and the separate term in the sum is an involved Laurent polynomial of $x$. This makes it difficult to work with this expression.

The Askey-Wilson polynomial (\ref{AWP}) is an eigenfunction of the Koornwinder Hamiltonian (\ref{KH}):
\be
H_1^K\ K_{[n]}(x)&=&\Lambda_n^K(q,a,b,c,d)\ K_{[n]}(x)\nn\\
\\
\Lambda_n^K(q,a,b,c,d)&=&\{\eta q^n\}\{q^n\}\nn
\ee
In the case of type $A$ root system, one would just put $n$ in (\ref{AWP}) to be non-integer and extend the sum up to infinity. It, however, does not work with the AW polynomials: at non-integer $n$ they are no longer eigenfunctions of the Hamiltonian (\ref{KH}). An eigenfunction with an arbitrary eigenvalue is more complicated, it is the Askey-Wilson function.

\subsubsection{AW functions}

There are various forms of the AW functions, which are defined to be eigenfunctions of the rank one Koornwinder Hamiltonians with arbitrary eigenvalues (universal solutions, in terms of \cite{dFK2}). One of the possible forms is the following one (see \cite{KoornM} for a review):
\be\label{Kf}
\mathcal{K}(x,y;q,a,b,c,d)\sim \Big[{q^2dy\over\eta}\Big/{q^2\over a};x\Big]_\infty\cdot
\sum_{k=0}^\infty \left({q\over\bar ay}\right)^k{1-\bar d\bar b\bar c yq^{4k-2}\over 1-\bar d\bar b\bar c yq^{-2}}
{(\bar d\bar b\bar c yq^{-2},\bar d y,\bar by,\bar cy;q^2)_k\over(q^2,\bar b\bar c,\bar d\bar b,\bar d\bar c;q^2)_k}
\Big[d\Big/{\bar d\bar b\bar c y\over d};x\Big]_k
\ee
where we use the notation
\be
[a/b;x]_k:={(ax,ax^{-1};q^2)_k\over(bx,bx^{-1};q^2)_k}
\ee
and similarly for $[...]_\infty$.
The function $\mathcal{K}(x,y;q,a,b,c,d)$ is an eigenfunction of the Koornwinder Hamiltonian with the eigenvalue
\be
\Lambda_y^K(q,a,b,c,d)=y+{1\over y}-\eta-{1\over\eta}=\langle y;\eta\rangle
\ee
In order to obtain the AW polynomial from this function, one rewrites \cite{KoSt} formula (\ref{Kf})
using Bailey's formula, in the form
\be\label{Kf2}
\mathcal{K}(x,y;q,a,b,c,d)\sim \sum_{k=0}^\infty{(dx,dx^{-1},\bar dy,\bar dy^{-1};q^2)_k\over(q^2,db,dc,da;q^2)_k}q^{2k}+
\ee
$$
+{(q^2ba^{-1},q^2ca^{-1},q^2a^{-1}d^{-1};q^2)_\infty\over (db,dc,daq^{-2};q^2)_\infty}\Big[\bar d\Big/{q^2\over\bar a};y\Big]_\infty
\Big[d\Big/{q^2\over a};x\Big]_\infty
\sum_{k=0}^\infty{(q^2a^{-1}x,q^2a^{-1}x^{-1},q^2\bar a^{-1}y,q^2\bar a^{-1}y^{-1};q^2)_k\over
(q^2,q^2ba^{-1},q^2ca^{-1},q^4a^{-1}d^{-1};q^2)_k}q^{2k}
$$
Now the AW polynomial is obtained from this formula upon the reduction $y=\bar d q^{2n}=\eta q^{2n}$, since the coefficient in front of the second sum in (\ref{Kf2}) vanishes at this reduction. Thus, an extension of (\ref{AWP}) to an eigenfunction with an arbitrary eigenvalue (aka the AW function) requires adding a new term.

\subsection{AW BA function\label{5.4}}

Already at the level of rank 1, one can see that the AW function is not looking to be an appropriate candidate for the universal solution, i.e. a counterpart of the NS function: it is symmetric under the Weyl group action $x\to x^{-1}$, while the BA function is not. Hence, one needs some other eigenfunction. In this subsection, we will construct the BA function directly and discuss the corresponding universal solution, which is not, however, the AW function.

\subsubsection{$B_1$ BA function\label{BBA}}

The key property of the AW case is that the eigenfunctions do not depend on $t$, and, hence, the corresponding BA function does not depend on $m$ at $t=q^{-m}$. At the same time, it depends on other $m_a$ associated with the Koornwinder parameters.

The values of parameters $a,b,c,d$ used in sec.\ref{KBA} for constructing the Koornwinder BA function do not correspond to reduced systems of roots (and to the corresponding BA functions), since some of $m_i$’s are equal to -1 in those cases. However, one can extend the
allowed region of $m_i$’s to these values: it is enough if $M_1$ and $M_2$ of sec.\ref{KBA} would be non-negative, and all the formulas persist. Moreover, in the cases describing the root
systems with roots of not more than two lengths, there are serious simplifications at integer values of $m_i$’s that
allow one to construct simpler formulas for the Baker-Akhiezer functions.

We start with considering the simplest example of the BA function, the set $(a,b,c,d)=(-1,-q,q^{-2m},q)$, $m\in\mathbb{Z}_>0$ (i.e. $M_1=0$, $M_2=m$) corresponding to the case of $B_1$ root system. In this case,
\be\label{BAB}
\Psi_m^{B_1}(x,y;q)= x^{\lambda+m\over 2} \sum_{k=0}^{m}\prod_{j=1}^k{[2(m-j+1]_q[2(j-\lambda-m-1)]_q\over[2j]_q[2(\lambda-j)]_q}x^{-k}
\ee
where $y:=q^{2\lambda}$. The simplest way to obtain this formula is to notice that the Koornwinder Hamiltonian in this case reduces to the Macdonald-Ruijsenaars Hamiltonian (\ref{MR}) plus a constant term. Another way to derive it is to use the periodicity condition (\ref{BAgeneq}), see Appendix A.

Upon the replacement $q^2\to q$, formula (\ref{BAB}) coincides with formula (\ref{BA2}) for the Baker-Akhiezer function for $A_1$:
\be
\Psi_m^{B_1}(x,y;q)=\overline{\Psi}_m^{A_1}(x,y;q^2)
\ee
We remind that, in the $C^\vee C$ case, we use parameter $q^2$ instead $q$, i.e. this identity is certainly expected,
and similarly below in (\ref{C1})-(\ref{C1d}).

One can generate the AW polynomial from this BA function by summing over the Weyl group, i.e. over the reflection:
\be
K_{[n]}(x;-1,-q,q^{-2m},q)=\Psi_m^{B_1}(x,q^{4n-2m};q)+\Psi_m^{B_1}(x^{-1},q^{4n-2m};q)
\ee

An important property of the BA function (\ref{BAB}) is that is an eigenfunction of the Koornwinder Hamiltonian:
\be
H_1^K\ \Psi_m^{B_1}(x,y;q)&=&\Lambda_y^{B_1}(q,a,b,c,d)\ \Psi_m^{B_1}(x,y;q)\nn\\
\\
\Lambda_y^{B_1}(q,a,b,c,d)&=&\{q^{\lambda-m\over 2}\}\{q^{\lambda+m\over 2}\}\nn
\ee
However, it is not symmetric under the Weyl group action $x\to x^{-1}$, which means that it is {\bf a different eigenfunction} of the Koornwinder Hamiltonian, which does not coincide with the AW function!

The properly normalized BA function,
\be
\Psi_m^{B_1}(x,y;q)=\prod_{j=1}^m\{q^{\lambda-j}\}\cdot
x^{\lambda+m\over 2} \sum_{k=0}^{m}\prod_{j=1}^k{[2(m-j+1]_q[2(j-\lambda-m-1)]_q\over[2j]_q[2(\lambda-j)]_q}x^{-k}
\ee
is symmetric under permutation of $x$ and $y$,
\be\label{Bsym}
\Psi_m^{B_1}(x,y;q)=\Psi_m^{B_1}(y,x;q)
\ee
Note that this property {\bf differs} from symmetricity in (\ref{Ksym}), where an additional changing to the dual AW parameters is implied, i.e. changing the system. Symmetric property (\ref{Bsym}) is within the $B$ type system.

Because of equivalence with the $A_1$ BA function, we know the universal solution in this $B_1$ case, i.e. $(a,b,c,d)=(-1,-q,t_s,q)$: it is just the power series
\be\label{NSB1}
\mathfrak{K}(x;q,t,-1,-q,t_s,q)=x^{\lambda\over 2}\ \sum_{k=0}^{\infty}\left({1\over x}\right)^{k+{1\over 4}\log_q t_s}
{\Big({q^2\over t_s}yq^{2k},t_s;q^2\Big)_{k}
\over \Big(yq^{2k},q^2;q^2\Big)_{k}}
\ee
which coincides, of course, with the $n=2$ NS function (\ref{NS2}) upon the substitution $q^2\to q$.

\subsubsection{BA function for other root system}

In the case of systems parameterized by one additional parameter $t_l$ or $t_s$, the formulas are particularly simple. In this paragraph we keep $m\in\mathbb{Z}_>0$.
Similarly to the case of $B_1$ root system, the set $(a,b,c,d)=(-q^{-m},-q^{1-m},q^{-m},q^{1-m})$ in the $C_1$ case gives rise to
\be\label{C1}
\Psi_m^{C_1}(x,y;q)=\Psi_m^{A_1}(x^2,y;q^2)
\ee
the set $(a,b,c,d)=(-q,-q^{-2m},q,q^{-2m})$ in the $B_1^\vee$ case, to
\be\label{B1d}
\Psi_m^{B_1^\vee}(x,y;q)=\Psi_m^{A_1}(x^2,y;q^4)
\ee
the set $(a,b,c,d)=(-1,-q^{1-2m},q^{2-2m},q)$ in the $C_1^\vee$ case, to
\be\label{C1d}
\Psi_m^{C_1^\vee}(x,y;q)=\Psi_{2m-1}^{A_1}(x,y^2;-q)
\ee
At last, the $D_1$ case is trivial, and such is the BA function:
\be
\Psi^{D_1}_m(y,x)=x^{\log_q(y)}
\ee
These BA functions are eigenfunctions of the corresponding Koornwinder Hamiltonians; choosing a proper normalization, one again obtains the BA functions symmetric under the permutation of $x$ and $y$; and using continuation (\ref{NS2}) to arbitrary values of parameters, one immediately arrives at the universal solution.

However, as we already pointed out, the universal solutions do not coincide with the AW functions being non-invariant w.r.t. to the Weyl group action $x\to x^{-1}$.

Some more details about the simplest BA functions for the root systems $B$ and $C$ can be found in Appendix A.

\subsubsection{On generic $C^\vee C_1$ BA function}

In the considered cases describing the root systems with roots of not more than two distinct lengths, there were serious simplifications at integer values of $m_i$'s that allowed us to construct simpler formulas for the Baker-Akhiezer functions: all the coefficients in the BA function could be presented in a factorized form. In other cases, this is impossible.

Indeed, looking at the generic $C^\vee C_1$ case, we just consider a polynomial BA
\be
\Psi_{\{m_i\}}(x,y;q)&=&x^{\lambda+m\over 2} \sum_{k=0}^{M_1+M_2}\psi_k(\lambda,\{m_i\})x^{-k}
\ee
that satisfies the periodicity condition (\ref{Kpc}). It generates the AW polynomial
\be
K_{[n]}(x;q,-q^{-m_1},-q^{-m_2},q^{-m_3},q^{-m_4})&=&\Psi_{\{m_i\}}(x,q^{4n-2M_1-2M_2};q)+\Psi_{\{m_i\}}(x^{-1},q^{4n-2M_1-2M_2};q)\nn\\
\ee
we use the notation $m_1+m_2=2M_1-1$, $m_3+m_4=2M_2-1$.

There is a symmetry under permutations $m_1\leftrightarrow m_2$ and $m_3\leftrightarrow m_4$. Then,
\be
\psi_0(\lambda,\{m_i\})&=&1\nn\\
\psi_1(\lambda,\{m_i\})&=&{[\lambda+M_1+M_2]_q
\over[2(\lambda-1)]_q}{(a+b+c+d+a^{-1}+b^{-1}+c^{-1}+d^{-1})\over (q^2-1)}\nn\\
&\ldots&
\ee
$\psi_{0,1}(\mu,\{m_i\})$ are exactly the coefficients of the AW polynomial: they do not simplify at integer values of $\{m_i\}$, just the coefficients at $k>M_1+M_2$ vanish. This is what happens with all coefficients: their complexity grows fast with level. Hence, in this generic AW case, an explicit formula for the universal solution remains unavailable.

\section{Conclusion}

In this paper, we reviewed the structure and properties of the symmetric and non-symmetric Koornwinder polynomials as compared with the corresponding Macdonald polynomials. We demonstrated that basically all the properties have Macdonald counterparts. There are only two essential differences with the Macdonald case, both have to do with generalizations of these polynomials.

\begin{itemize}

\item The first problem has something to do with the symmetric Koornwinder polynomials: it is not simple to construct a universal solution, which is an eigenfunction of the Koornwinder Hamiltonians (sec.\ref{5}) with arbitrary eigenvalues, and is a counterpart of the NS function. In fact, the NS function was constructed \cite{NS} basing on the branching rule (and then using the nested anzatz \cite{NS,MMP7}), which, in the case of Macdonald polynomials, has a structure convenient for the continuation to arbitrary complex eigenvalues. However, as we explain in sec.\ref{secbr}, this is not the case for the symmetric Koornwinder polynomials. Moreover, it looks so that a possible continuation would imply emerging new structures not seen in the polynomial case.

    In fact, as we explain in secs.\ref{5.3}-\ref{5.4}, even in the $n=1$ (Askey-Wilson) case, there are problems with continuation, though at peculiar values of parameters the universal solution is available because of coincidence with the $A_1$ root system.

\item The second problem is that it is not looking possible to construct a counterpart of twisting \cite{CE,CF,MMP1,NSM1,NSM2,NSM4} in the Koornwinder case. It is related with the lack of automorphisms of DAHA in the case of root systems of non-$A$ type.

\end{itemize}

Let us note that we avoided a discussion of algebraic role of the symmetric Koornwinder polynomials within the context of the $n$-body representation of an algebra which is obtained from the DIM algebra by inserting the orientifold plane (projecting to the reflection states in the spirit of \cite{Matsuo}, see also algebra $K$ in \cite[sec.4]{FJMV}). We discuss these matters elsewhere.

Another interesting direction of generalizations is possible elliptization(s) of the symmetric Koornwinder polynomials similarly to what is done for the symmetric Macdonald polynomials \cite{Shi,FOS,AKMM2,MMP4,MMP5}. It will be also discussed elsewhere.

\section*{Acknowledgements}

We are grateful to M. Matushko, S. Mironov, I. Ryzhkov, N. Tselousov and Y. Zenkevich for useful discussions and explanations. This work is supported by RSF grant 26-12-00191.

\section*{Appendix A. BA functions for concrete root systems}

In this paper, we considered root systems that contain the following types of roots: $\varepsilon_i$, $\varepsilon_i-\varepsilon_k$, $\varepsilon_i+\varepsilon_k$ and $2\varepsilon_i$. Let us list the periodicity equations (\ref{BAgeneq}) in terms of manifest coordinates $\vec x$ for these roots:
\begin{align}
    \alpha = \varepsilon_i-\varepsilon_k &\quad \quad \Psi_{\{m_{\alpha}\}}(x_i q^j,\vec{y};\,q) = \Psi_{\{m_{\alpha}\}}(x_k q^j,\vec{y};\,q) \quad &&\text{at}\quad x_i = x_k, \\
     \alpha = \varepsilon_i+\varepsilon_k &\quad \quad \Psi_{\{m_{\alpha}\}}(x_i q^j,\vec{y};\,q) = \Psi_{\{m_{\alpha}\}}(x_k q^{-j},\vec{y};\,q) \quad &&\text{at}\quad x_i = x_k^{-1},\\
    \alpha = \varepsilon_i &\quad \quad \Psi_{\{m_{\alpha}\}}(x_i q^j,\vec{y};\,q) = \Psi_{\{m_{\alpha}\}}(x_i q^{-j},\vec{y};\,q) \quad &&\text{at}\quad x_i = 1,\\
    \alpha = 2\varepsilon_i &\quad \quad \Psi_{\{m_{\alpha}\}}(x_i q^{j/2},\vec{y};\,q) = \Psi_{\{m_{\alpha}\}}(x_i q^{-j/2},\vec{y};\,q) \quad &&\text{at}\quad x_i^2 = 1,
\end{align}

\paragraph{BA function for root system $B_1$.}
There is only one root $\alpha = \varepsilon_1$ and $\alpha^{\vee} = 2\varepsilon_1$. We associate the parameter $m_s$ with this root, $\rho_{m_s} = 1/2m_s$. The BA function
\begin{equation}
    \Psi_{m_s}^{B_1} = q^{(\lambda +1/2 m_s)z} \sum_{l=0}^{m_s} \psi_l x^{-l} = x^{\lambda +1/2 m_s} \sum_{l=0}^{m_s} \psi_l x^{-l}
\end{equation}
satisfies the periodicity equations
\begin{equation}
    \Psi_{m_s}^{B_1}\left(x q^{j}\right) =  \Psi_{m_s}^{B_1}\left(x q^{-j}\right) \quad \text{at} \quad x = 1
\end{equation}
for $j = 1, \dots, m_s$
\begin{equation}
    \sum_{l=0}^{m_s} \psi_l\left (q^{(2\lambda+m_s)j}\frac{1}{q^{jl}} -q^{jl} \right) = 0 \quad \text{for} \quad j = 1, \dots, m_s
\end{equation}

\paragraph{BA function for root system $C_1$.}
There is only one root $\alpha = 2\varepsilon_1$ and $\alpha^{\vee} = \varepsilon_1$. We associate the parameter $m_l$ with this root, $\rho_{m_l} = m_l$. The BA function
\begin{equation}
    \Psi_{m_l}^{C_1} = q^{(\lambda + m_l)z} \sum_{l=0}^{m_l} \psi_l x^{-2l} = x^{\lambda +m_l} \sum_{l=0}^{m_l} \psi_l x^{-2l}
\end{equation}
satisfies the periodicity equations
\begin{equation}
    \Psi_{m_l}^{C_1}\left(x q^{j/2}\right) =  \Psi_{m_l}^{C_1}\left(x q^{-j/2}\right) \quad \text{at} \quad x^2 = 1
\end{equation}
for $j = 1, \dots, m_l$
\begin{equation}
    \sum_{l=0}^{m_l} \psi_l\left (q^{(\lambda+m_s)j}\frac{1}{q^{jl}} -q^{jl} \right) = 0 \quad \text{for} \quad j = 1, \dots, m_l
\end{equation}

\paragraph{BA function for root system $B_2$.}

The roots are $\varepsilon_1-\varepsilon_2$, $\varepsilon_2$, $\varepsilon_1+\varepsilon_2$, $\varepsilon_1$. Roots $\varepsilon_1\pm\varepsilon_2$ are associated with $m$ and $\varepsilon_i$ are associated with $m_s$, $\rho_{m,m_s} = \left(m+m_s/2,m_s/2 \right)$, the BA function is
\begin{equation}
\begin{aligned}
    \Psi_{m,m_s}^{B_2} = x_1^{\lambda_1+m+m_s/2}x_2^{\lambda_2+m_s/2} \sum_{l_1,l_2,l_3,l_4=0}^{m_l} \psi_{l_1,l_2,l_3,l_4} \left(\frac{x_2}{x_1}\right)^{l_1} \left(\frac{1}{x_2}\right)^{l_2}\left(\frac{1}{x_1 x_2}\right)^{l_3} \left(\frac{1}{x_1}\right)^{l_1} = \\x_1^{\lambda_1+m+m_s}x_2^{\lambda_2+m_s} \sum_{l_1,l_2,l_3,l_4=0}^{m_l} \psi_{l_1,l_2,l_3,l_4}\,\, x_1^{-(l_1+l_3+l_4)} x_2^{l_1-l_2-l_3}
\end{aligned}
\end{equation}
and the periodicity equations are
\begin{align}
   & \Psi_{m,m_s}^{B_2} \left(x_1 q^{j} \right) = \Psi_{m,m_s}^{B_2} \left(x_2 q^{j} \right) \quad \text{at} \quad x_1 = x_2,\\
   &  \Psi_{m,m_s}^{B_2} \left(x_2 q^{j} \right) = \Psi_{m,m_s}^{B_2} \left(x_2 q^{-j} \right) \quad \text{at} \quad x_2 = 1,\\
    & \Psi_{m,m_s}^{B_2} \left(x_1 q^{j} \right) = \Psi_{m,m_s}^{B_2} \left(x_2 q^{-j} \right) \quad \text{at} \quad x_1 = x_2^{-1}, \\
    &  \Psi_{m,m_s}^{B_2} \left(x_1 q^{j} \right) = \Psi_{m,m_s}^{B_2} \left(x_1 q^{-j} \right) \quad \text{at} \quad x_1 = 1,
\end{align}

\paragraph{BA function for root system $C_2$.}

The roots are $\varepsilon_1-\varepsilon_2$, $2\varepsilon_2$, $\varepsilon_1+\varepsilon_2$, $2\varepsilon_1$. Roots $\varepsilon_1\pm\varepsilon_2$ are associated with $m$ and $2\varepsilon_i$ are associated with $m_l$, $\rho_{m,m_l} = \left(m+m_l,m_l \right)$, the BA function is
\begin{equation}
\begin{aligned}
    \Psi_{m,m_l}^{C_2} = x_1^{\lambda_1+m+m_l}x_2^{\lambda_2+m_l} \sum_{l_1,l_2,l_3,l_4=0}^{m_l} \psi_{l_1,l_2,l_3,l_4} \left(\frac{x_2}{x_1}\right)^{l_1} \left(\frac{1}{x_2^2}\right)^{l_2}\left(\frac{1}{x_1 x_2}\right)^{l_3} \left(\frac{1}{x_1^2}\right)^{l_1} = \\x_1^{\lambda_1+m+m_l}x_2^{\lambda_2+m_l} \sum_{l_1,l_2,l_3,l_4=0}^{m_l} \psi_{l_1,l_2,l_3,l_4}\,\, x_1^{-(l_1+l_3+2l_4)} x_2^{l_1-2l_2-l_3}
\end{aligned}
\end{equation}
and the periodicity equations are
\begin{align}
   & \Psi_{m,m_l}^{C_2} \left(x_1 q^{j} \right) = \Psi_{m,m_l}^{C_2} \left(x_2 q^{j} \right) \quad \text{at} \quad x_1 = x_2,\\
   &  \Psi_{m,m_l}^{C_2} \left(x_2 q^{j/2} \right) = \Psi_{m,m_l}^{C_2} \left(x_2 q^{-j/2} \right) \quad \text{at} \quad x_2^2 = 1,\\
    &  \Psi_{m,m_l}^{C_2} \left(x_1 q^{j} \right) = \Psi_{m,m_l}^{C_2} \left(x_2 q^{-j} \right) \quad \text{at} \quad x_1 = x_2^{-1}, \\
    &  \Psi_{m,m_l}^{C_2} \left(x_1 q^{j/2} \right) = \Psi_{m,m_l}^{C_2} \left(x_1 q^{-j/2} \right) \quad \text{at} \quad x_1^2 = 1,
\end{align}

\section*{Appendix B. Non-symmetric Koornwinder polynomials: adding zero parts to labels}

Here we consider a few simple examples of non-symmetric Koornwinder polynomials $E^C_\alpha$ with arbitrary $n$ and $\alpha$ containing a fixed set of non-zero entries, while the number of zero entries increases with $n$. The first example of such $\alpha$ is given by a set of the form
\begin{align}
  \vec\alpha_{LR} = [\underbrace{0,\dots,0}_L, \vec\alpha, \underbrace{0,\dots,0}_R]
\end{align}
It turns out to be possible to obtain universal formulas, at least for first few small
$\vec\alpha$, and we list them here
\newcommand{\valr}[0]{
  \vec\alpha_{LR}
}
\newcommand{\lyr}[1]{
  [#1]_{LR}
}
\subsubsection*{Level 1}
Simplest is the case $\vec\alpha = [-1]$. The universal answer is
\begin{align}
  E^C_{\lyr{-1}} =
  \frac{1}{x_{L+1}}
  + \frac{(t-1)(t+1) t^{-2} Q_L^2 Q_R^4 a b c d}{\left(a b c d Q_L^2 Q_R^4 - 1\right)}
    p_{-1,L}
\end{align}
here $x_{L+1}$ is situated precisely on the spot where $-1$ is in $\lyr{-1}$,
$Q_L = t^L$, $Q_R = t^R$ encode the dependence on length of the diagram $n$.
\footnote{The observation is that
this dependence of non-symmetric Koornwinder polynomials is only through $Q_L$
and $Q_R$ and not, say, through $t^{L^2}$.}
Define the left (right) power sums as sums over variables $x_i$
with $i$ to the left (right) of $\valr$:
\begin{align}
  p_{k,L} = & \ \sum_{i=1}^L x_i^k
  \\ \notag
  p_{k,R} = & \ \sum_{i=1}^{R} x_{i + L+l_\alpha}^k
\end{align}

The other polynomial at this level is only slightly more complicated but is expressed in the same terms
\begin{align}
  E^C_{\lyr{1}} = & \ x_{L+1}
  \\ \notag
  + & \ \frac{1}{x_{L+1}}
  \cdot
  \frac{Q_L^2
    \left(
    a b c d (q^4 (t^2 -1) Q_L^4 Q_R^4 + q^4 Q_L^4 Q_R^2 - q^2 Q_L^2 Q_R^2)
    + a c (q^4 t^2 Q_L^4 Q_R^2 - q^2 t^2 Q_L^2 Q_R^2 - q^2 Q_L^2 + 1)
    (- q^4 t^2 Q_L^2 + q^2)
    \right)
  }{(a b c d  q^2 Q_L^4 Q_R^4 - 1) (q^2 t^2 Q_L^2-1)}
  \\ \notag
  + & \
  \frac{q^2 Q_L^2 (t^2 - 1)}{(q^2 t^2 Q_L^2- 1)}
  \left(p_{1,R} + p_{-1,R}\right)
  +
  \frac{a c (t-1)(t+1) t^{-2} Q_L^2 \left(b d q^2 Q_L^2 Q_R^2 - 1\right)}
       {(a b c d q^2 Q_L^4 Q_R^4-1)}
  \frac{(q^2 t^2 Q_L^2 Q_R^2 - 1)}{(q^2 t^2 Q_L^2 - 1)}
  p_{-1,L} \\ \notag
  - & \ \frac{ Q_L^2 (q^2 t^2 Q_L^2 Q_R^2 - 1)
    \left(a b c d \left(\frac{1}{a}+\frac{1}{b}+\frac{1}{c}+ \frac{1}{d}\right)
     q^2 Q_L^2 Q_R^2 - (b + d)q^2 - (a+c) \right)}
  {(a b c d q^2 Q_L^4 Q_R^4 - 1) (q^2 t^2 Q_L^2 - 1)}
\end{align}

Even at this starting point one may already wonder:
\begin{itemize}
\item Which structures in variables $x$ occur in the formula for particular $E^C_{\valr}$?
  \item Is it possible to predict/limit the shape of the coefficients in front of $x$-structures?
At the very least, is it possible to estimate the limits of the degrees of $Q_{L,R}$ in the numerators and
denominators?
\end{itemize}

We continue to the next level.

\subsubsection*{Level 2}

At level 2, there are two distinct possibilities: the partition may have one
or two parts. Therefore, one may study (in addition to $L$- and $R$ -dependencies) how the two-part formula behaves with additional $M$ zeroes between the parts.

In particular,
$[-1,-1]_{LMR} = [\underbrace{0,\dots,0}_{L},-1,\underbrace{0,\dots,0}_{M},-1,\underbrace{0,\dots,0}_{R}]$ and

\newcommand\efm[0]{
  e_4^{-1}
}

\begin{align}
  E^C_{[-1,-1]_{LMR}} = & \
  x_{L+1}^{-1} x_{L+M+2}^{-1}
  + p_{-1,L} x_{L+1}^{-1} \cdot
  \frac{(1-t^{-2})
    (t^2 - \efm Q_L^{-2} Q_M^{-4} Q_R^{-4})}
       {(t^2 - \efm Q_L^{-2} Q_M^{-2} Q_R^{-4})
         (1 - \efm Q_L^{-2} Q_M^{-4} Q_R^{-4})}
       \\ \notag
       + & \ x_{L+1}^{-1} p_{-1,M} \cdot
       \frac{(t^2-1)}
            {(t^2 - \efm Q_L^{-2} Q_M^{-2} Q_R^{-4})}
            + p_{-1,L} x_{L+M+2}^{-1}
            \frac{(1 - t^{-2})
              (t^2 - \efm Q_L^{-2} Q_M^{-2} Q_R^{-4})}
                 {(t^2 -\efm Q_L^{-2} Q_M^{-2} Q_R^{-4})
                   (1- Q_L^{-2} Q_M^{-4} Q_R^{-4})}
                 \\ \notag
                 + & \
                 \left(\frac{p_{-1,L}^2}{2} - \frac{p_{-2,L}}{2}\right)\cdot
                 \frac{(1 -t^{-4})
                 }{(t^2 - \efm Q_L^{-2} Q_M^{-2} Q_R^{-4})(1- \efm Q_L^{-2} Q_M^{-4} Q_R^{-4} - 1)}
                 \\ \notag
                 + & \
                 p_{-1,L}p_{-1,M}\cdot
                 \frac{(1-t^{-2})t^2}{(t^2 - \efm Q_L^{-2} Q_M^{-2} Q_R^{-4})
                   (1 - \efm Q_L^{-2} Q_M^{-4} Q_R^{-4})}
                 \\ \notag
                 - & \ p_{-1,L}
                 \cdot
                 \frac{
                   (1 - t^{-4})t^2 \efm
                   Q_L^2 Q_M^4 Q_R^6
                   \left(
                   Q_L^2 Q_M^2 Q_R^2 (a b d + b c d) - b - d
                   \right)}{
                   (t^2 Q_L^2 Q_M^2 Q_R^4 - \efm)
                   (Q_L^2 Q_M^4 Q_R^4 - \efm)}
                 \\ \notag
                 - & \ p_{-1,M}\cdot
                 \frac{(t^2-1) \efm
                   Q_L^2 Q_M^4 Q_R^6
                   \left(
                   Q_L^2 Q_M^2 Q_R^2 (a b d + b c d) - b - d
                   \right)}
                 {(t^2 Q_L^2 Q_M^2 Q_R^4 - \efm)
                   (Q_L^2 Q_M^4 Q_R^4 - \efm)}
                 \\ \notag
                 - & \ x_{L+1}^{-1}\cdot
                 \frac{Q_R^2 \efm\left(
                   Q_L^2 Q_M^2 Q_R^2 (a b d + b c d) - b - d
                   \right)
                   (t^2 Q_L^2 Q_M^4 Q_R^4 - \efm)
                 }{(t^2 Q_L^2 Q_M^2 Q_R^4 - \efm)
                   (Q_L^2 Q_M^4 Q_R^4 - \efm)}
                 \\ \notag
                 - & \ x_{L+M+2}^{-1} \cdot
                 \frac{Q_M^2 Q_R^2 \efm\left(
                   Q_L^2 Q_M^2 Q_R^2 (a b d + b c d) - b - d
                   \right)}{(Q_L^2 Q_M^4 Q_R^4 - \efm)}
                 \\ \notag
  - & \frac{\scriptscriptstyle Q_M^2 Q_R^4
    \left(Q_L^4 Q_M^4 Q_R^4 (a^2 b^2 d^2 t^2 + a b^2 c d^2 (t^2+1) + b^2 c^2 d^2 t^2)
    - Q_L^2 Q_M^2 Q_R^2 (a b^2 d (t^2+1) + a b d^2 (t^2+1)+ b^2 c d (t^2+1)
    + b c d^2 (t^2+1))
    + (b^2 + b d (t^2+1) + d^2)\right)}
  {(e_4 t^2 Q_L^2 Q_M^2 Q_R^4 - 1)(e_4 Q_L^2 Q_M^4 Q_R^4 - 1)}
\end{align}

One can see that all degree two terms are completely ``Pochhammerized'' (factorized
into Pochhammer-like expressions) and most complicated is the free (degree zero) term. Whether these partly Pochhammerized structures can be additionally simplified by a smart choice of non-homogeneous basis elements {\it a la} (\ref{111})-(\ref{112}) remains to be seen. This would definitely help  in resolving the mystery of the Koornwinder universal solution.


\begin{thebibliography}{12}

\bibitem{MMMP1} A.~Mironov, V.~Mishnyakov, A.~Morozov, A.~Popolitov,
JHEP, \textbf{23} (2020) 065,
arXiv:2306.06623

\bibitem{Pope} C.N. Pope, L.J. Romans and X. Shen, Phys.Lett. {\bf B236} (1989) 173-178; Nucl.Phys. {\bf 339B} (1990) 191-221; Phys.Lett. {\bf B242} (1990) 401-406; Phys.Lett. {\bf B245} (1990) 72-78

\bibitem{Awata} H.~Awata, M.~Fukuma, Y.~Matsuo and S.~Odake,
Prog. Theor. Phys. Suppl. \textbf{118} (1995) 343-374,
hep-th/9408158

\bibitem{KR1} V.G. Kac and A. Radul, Comm.Math.Phys. {\bf 157} (1993) 429-457, hep-th/9308153

\bibitem{MMMP2} A.~Mironov, V.~Mishnyakov, A.~Morozov, A.~Popolitov,
Phys. Lett. \textbf{B845} (2023) 138122,
arXiv:2307.01048

\bibitem{Tsy} A.~Tsymbaliuk,
Adv. Math. \textbf{304} (2017) 583-645,
arXiv:1404.5240

\bibitem{Proch}  T.~Proch\'azka,
JHEP, \textbf{10} (2016) 077,
arXiv:1512.07178

\bibitem{MMP} A. Mironov, A. Morozov, A. Popolitov, JHEP, \textbf{09} (2024) 200,
  arXiv:2406.16688

\bibitem{DI} J. Ding, K. Iohara, 
Lett. Math. Phys. {\bf 41} (1997) 181-193, q-alg/9608002

\bibitem{Miki} K. Miki, J. Math. Phys. {\bf 48} (2007) 123520

\bibitem{K} M. Kapranov,  
Algebraic geometry {\bf 7}, J.Math. Sci. {\bf 84} (1997) 1311-1360, alg-geom/9604018

\bibitem{BS} I. Burban, O. Schiffmann, 
Duke Math. J. {\bf 161} (2012) 1171, arXiv:math/0505148

\bibitem{S}  O. Schiffmann, 
J. Algebraic Combin. {\bf 35} (2012) 237-26, arXiv:1004.2575

\bibitem{Feigin} B. Feigin, M. Jimbo, T. Miwa, E. Mukhin,
Commun. Math. Phys. \textbf{356}  (2017) 285, arXiv:1603.02765

\bibitem{RS} S.N.M. Ruijsenaars, H. Schneider, Ann.Phys. (NY), {\bf 170} (1986) 370\\
S.N.M. Ruijsenaars, Comm.Math.Phys. {\bf 110} (1987) 191-213

\bibitem{Miki1} K. Miki, 
Lett. Math. Phys. {\bf 47} (1999) 365-378

\bibitem{dFK1} P. Di Francesco, R. Kedem, Comm. Math. Phys. {\bf 369(3)} (2019) 867-928, arXiv:1704.00154

\bibitem{book:Ch-daha} I. Cherednik, {\sl Double affine Hecke algebras},
  Vol. {\bf 319}, Cambridge University Press, 2005

\bibitem{NSM4} A.~Mironov, A.~Morozov and A.~Popolitov,
Phys. Lett. \textbf{B879} (2026) 140592,
arXiv:2602.21120

\bibitem{NS} M. Noumi, J. Shiraishi, 
arXiv:1206.5364

\bibitem{Cha} O. Chalykh,  Adv.Math. {\bf 166(2)} (2002) 193-259, math/0212313

\bibitem{dFK2} P.~Di Francesco, R.~Kedem,
Selecta Math. \textbf{30} (2024) no.2, 23,
arXiv:2112.09798

\bibitem{MMP3} A.~Mironov, A.~Morozov, A.~Popolitov,
Phys. Lett. \textbf{B869} (2025) 139840,
arXiv:2411.16517

\bibitem{CE}   O. Chalykh, P. Etingof, Advances in Mathematics, {\bf 238} (2013) 246-289,
arXiv:1111.0515

\bibitem{CF} O. Chalykh, M. Fairon, 
J.Geom.Phys. {\bf 121} (2017) 413-437,
  arXiv:1704.05814

\bibitem{MMP1} A.~Mironov, A.~Morozov, A.~Popolitov,
Phys. Lett. \textbf{B863} (2025) 139380,
arXiv:2410.10685

\bibitem{NSM5} A.~Mironov, A.~Morozov, A.~Popolitov, to appear

\bibitem{Opd95} E.M. Opdam, 
Acta Mathematica, {\bf 175(1)} (1995) 75–121

\bibitem{Mac96} I.G. Macdonald, 
Asterisque-Societe Mathematique de France, {\bf 237} (1996) 189-208

\bibitem{Che95} I. Cherednik, 
IMRN, {\bf 1995(10)} (1995) 483, q-alg/9505029


\bibitem{NSM1} A.~Mironov, A.~Morozov, A.~Popolitov,
arXiv:2512.24811

\bibitem{NSM2} A.~Mironov, A.~Morozov, A.~Popolitov,
Nucl. Phys. \textbf{B1028} (2026) 117513,
arXiv:2601.10500

\bibitem{NSM3} A.~Mironov, A.~Morozov, A.~Popolitov,
Phys. Lett. \textbf{B877} (2026) 140457,
arXiv:2601.19878

\bibitem{Macr} I.G. Macdonald, 
Invent. Math. 15 (1972) 91–143

\bibitem{Mac} I.G. Macdonald, 
S\`eminaire Lotharingien Combin. {\bf 45} (2000), Article B45a, 40 pp, arXiv:math/0011046

\bibitem{MacConj} I.G. Macdonald, 
SIAM J.Math. Anal. {\bf 13:6} (1982) 988-1007

\bibitem{CherednikConj} I. Cherednik, 
Inventiones mathematicae, {\bf 125} (1996) 391, q-alg/9412016

\bibitem{CherednikDAHA} I. Cherednik, 
The Annals of Mathematics, Second Series, {\bf 141} (1995) 191-216

\bibitem{Koorn} T.H. Koornwinder, {\sl Askey-Wilson polynomials for root systems of type BC}, in: “Hypergeometric
functions on domains of positivity, Jack polynomials, and applications” (Tampa, FL, 1991),
Contemp. Math. {\bf 138} (1992) 189–204

\bibitem{vD} J.F.~van Diejen, 
Comp. Math. {\bf 95} (1995) 183-233, funct-an/9306002

\bibitem{HHL} J. Haglund, M. Haiman, N. Loehr, Am.J.Math. {\bf 130(2)} (2008) 359-383, math/0601693

\bibitem{KS} F. Knop, S. Sahi, 
Invent. Math. {\bf 128} (1997) 9-22, q-alg/9610016

\bibitem{BF} T.H. Baker, P.J. Forrester, 
q-alg/9701039

\bibitem{CO} K. Mimachi, M. Noumi, 
Duke Math. J. {\bf 95(1)} (1998) 621-634, q-alg/9610014

\bibitem{Macbook} I.G. Macdonald, {\it Symmetric functions and Hall polynomials},   Oxford University Press, 1995

\bibitem{CheIP} I. Cherednik, 
Sel. math., New ser. {\bf 3} (1997) 459–495, q-alg/9605014


\bibitem{CheF}  I. Cherednik, 
Invent. Math. {\bf 152} (2003) 213–303, math/0110024

\bibitem{Rdu} S.N. Ruijsenaars, Comm. Math. Phys., 115 (1988) 127-165

\bibitem{Etin}  P. Etingof, A. Varchenko, 
Duke Math. J. {\bf 104} (2000) 391-432. , math/9907181\\
P. Etingof, O. Schiffmann, A. Varchenko, 
Lett. Math. Phys. {\bf 62} (2002) 143-158\\
G. Felder, Y. Markov, V. Tarasov, A. Varchenko, 
Mathematical Physics, Analysis and Geometry, {\bf 3} (2000) 139-177, math/0001184\\
V. Tarasov, A. Varchenko, 
IMRN, {\bf 2000(15)} 801-829, math/0002132

\bibitem{MMZ} A.~Mironov, A.~Morozov, Y.~Zenkevich,
Eur. Phys. J. \textbf{C81} (2021) 461,
arXiv:2103.02508

\bibitem{MMdell} A.~Mironov, A.~Morozov,
Nucl. Phys. \textbf{B999} (2024) 116448,
arXiv:2309.06403

\bibitem{CheCMM}  I. Cherednik, IMRN, {\bf 1997 (10)} (1997) 449-467,
  q-alg/9702022

\bibitem{EK}  P. Etingof, A. Kirillov, Electr.Res.Announc.Amer.Math.Soc. {\bf 4} (1998) 43-47,
q-alg/9712051

\bibitem{MMPU} A.~Mironov, A.~Morozov, A.~Popolitov,
Phys. Rev. \textbf{D110} (2024) 126026,
arXiv:2410.03175

\bibitem{MMP7} A.~Mironov, A.~Morozov, A.~Popolitov,
Phys. Rev. \textbf{D113} (2026) no.12, 126019,
arXiv:2601.17453

\bibitem{Noumi} M. Noumi, Macdonald-Koornwinder polynomials and affine Hecke rings, Surikaisekikenkyusho Kokyuroku 919 (1995), pp. 44–55 (in japanese)

\bibitem{Sahi}  S. Sahi, 
Ann. Math. {\bf 150} (1999) 267–282, q-alg/9710032

\bibitem{Sahi2} S. Sahi, Some properties of Koornwinder polynomials, Contemp. Math. 254 (2000) 395-411

\bibitem{St}  J. Stokman, 
IMRN 19 (2000), 1005–1042, math/0002090

\bibitem{Chalykh} O. Chalykh, Commun. Math. Phys. {\bf 369(1)} (2019) 261-316, arXiv:1804.01766

\bibitem{CR} L. Colmenarejo, A. Ram, 
The Quarterly Journal of Mathematics, {\bf 76(3)} (2025) 983-1031, arXiv:2410.19957

\bibitem{AW} R. Askey, J. Wilson, 
Mem. Amer. Math. Soc. vol.{\bf 319}, 1985

\bibitem{Mimachi} K. Mimachi, 
Duke Math. J. {\bf 107(2)} (2001) 265–281

\bibitem{vDE} J.F. van Diejen, E. Emsiz, 
Journal of Algebra, {\bf 444} (20145) 606-614, arXiv:1408.2280

\bibitem{KoornM} T. Koornwinder, M. Mazzocco, 
 Indagationes Mathematicae (2025), arXiv:2407.17366

\bibitem{KoSt}  E. Koelink, J.V. Stokman, 
IMRN, {\bf 22} (2001) 1203–1227, math/0004053

\bibitem{Matsuo} J.-E. Bourgine, M. Fukuda, Y. Matsuo, R.-D. Zhu, JHEP, 2017 (2017) 15, arXiv:1709.01954

\bibitem{FJMV} B. Feigin, M. Jimbo, E. Mukhin, I. Vilkoviskiy, 
Selecta Mathematica, {\bf 27(4)} (2021) 52, arXiv:2003.04234

\bibitem{Shi} J. Shiraishi, 
J.Integrable.Syst. {\bf 4} (2019) xyz010, arXiv:1903.07495

\bibitem{FOS} M. Fukuda, Y. Ohkubo, J. Shiraishi, 
    SIGMA {\bf 16} (2020) 116, arXiv:2002.00243

\bibitem{AKMM2} H.~Awata, H.~Kanno, A.~Mironov, A.~Morozov,
JHEP, \textbf{08} (2020) 150,
arXiv:2005.10563

\bibitem{MMP4} A.~Mironov, A.~Morozov, A.~Popolitov, Z.~Zakirova,
Phys. Lett. B \textbf{865} (2025) 139467,
arXiv:2412.19588

\bibitem{MMP5} A.~Mironov, A.~Morozov, A.~Popolitov, Z.~Zakirova,
Pisma Zh. Eksp. Teor. Fiz. \textbf{121} (2025) no.9, 788-795,
arXiv:2503.07592

\end{thebibliography}
\end{document}